\documentclass[aps,prd,amsmath,amsfonts,amssymb,eqsecnum,nofootinbib,
  floatfix,secnumarabic,preprintnumbers,superscriptaddress,nobalancelastpage,
  onecolumn,notitlepage]{revtex4}
\usepackage[utf8]{inputenc}
\usepackage{color,graphicx,hyperref,dsfont,slashed,multirow}
\hypersetup{bookmarks=true,unicode=true,pdftoolbar=true,pdfmenubar=true,
  pdffitwindow=false,pdfstartview={FitH},pdftitle={UMSSM},
  pdfauthor={J.~Araz,~M.~Frank,~B.~Fuks}, pdfsubject={Subject},
  pdfcreator={Creator},pdfproducer={Producer}, pdfkeywords={Non-minimal 
  supersymmetry}{Beyond the Standard Model Physics}{sneutrino dark matter},
  pdfnewwindow=true,colorlinks=true,
  linkcolor=blue,citecolor=magenta,filecolor=magenta,urlcolor=cyan}
\newcommand{\be}{\begin{equation}}
\newcommand{\ee}{\end{equation}}
\def\bsp#1\esp{\begin{split}#1\end{split}}

\begin{document}
\date{\today}
\preprint{CUMQ/HEP 194}

\title{Differentiating $U(1)^\prime$ supersymmetric models with right
  sneutrino and neutralino dark matter}

\author{Jack Y. Araz}
\email{jack.araz@concordia.ca}
\affiliation{Concordia University 7141 Sherbrooke St. West, Montreal, QC,
 CANADA H4B 1R6}

\author{Mariana Frank}
\email{mariana.frank@concordia.ca}
\affiliation{Concordia University 7141 Sherbrooke St. West, Montreal, QC,
 CANADA H4B 1R6}

\author{Benjamin Fuks}
\email{fuks@lpthe.jussieu.fr}
\affiliation{Sorbonne Universit\'es, Universit\'e Pierre et Marie Curie
  (Paris 06), UMR 7589, LPTHE, F-75005 Paris, France}
\affiliation{CNRS, UMR 7589, LPTHE, F-75005, Paris, France}
\affiliation{Institut Universitaire de France, 103 boulevard Saint-Michel,
  75005 Paris, France}

\vspace{10pt}
\begin{abstract}
We perform a detailed analysis of dark matter signals of supersymmetric models
containing an extra $U(1)^\prime$ gauge group. We investigate scenarios in which
either the right sneutrino or the lightest neutralino are phenomenologically
acceptable dark matter candidates and we explore the parameter spaces of
different supersymmetric realisations featuring an extra $U(1)^\prime$. We
impose consistency with low energy observables, with known mass
limits for the superpartners and $Z^\prime$ bosons, as well as with Higgs boson
signal strengths, and we moreover verify that predictions for the anomalous
magnetic moment of the muon agree with the experimental value and require that
the dark matter candidate satisfies the observed relic density and direct and
indirect dark matter detection constraints. For the case where the sneutrino is
the dark matter candidate, we find distinguishing characteristics among
different $U(1)^\prime$ mixing angles. If the neutralino is the lightest
supersymmetric particle, its mass is heavier than that of the light sneutrino in
scenarios where the latter is a dark matter candidate, the parameter space is
less restricted and differentiation between models is more difficult. We finally
comment on the possible collider tests of these models.
\end{abstract}

\keywords{Dark Matter, UMSSM}

\maketitle

\section{Introduction}\label{sec:intro}

Supersymmetry is one of the most attractive theories of physics beyond the
Standard Model (SM). It introduces a viable space-time extension, provides a
natural solution to the hierarchy problem, allows for gauge coupling unification
at a single Grand Unified scale, and, last but not least, it predicts a stable,
neutral lightest supersymmetric particle (LSP) as a  realistic weakly
interacting massive particle  dark matter (DM) candidate. But despite the
numerous appealing aspects, low-energy supersymmetry (SUSY) is plagued by one
overwhelming failure: no compelling evidence for it is seen at the LHC. This
imposes stringent constraints on the masses of any supersymmetric coloured
particle. Under simplified assumptions, gluino and first and second
generation squark masses of less than 2~TeV are for instance excluded for a
large variety of LSP masses~\cite{ATLAS:2016uzr,ATLAS:2016lsr,ATLAS:2016kts}. 
The absence of any light superpartners so far hence puts the theory in serious
conflict with electroweak naturalness~\cite{Dine:2015xga,Hall:2011aa}. However,
most searches are based on the minimal supersymmetric scenario whose parameter
space left to explore at the
LHC is rapidly shrinking. In addition, the minimal model suffers from serious fine-tuning problems induced by the discovery of ATLAS~\cite{Aad:2012tfa} and CMS~\cite{Chatrchyan:2012xdj} collaborations of a scalar particle with a mass of 125 GeV and with the expected properties of a Standard Model Higgs boson. On one hand, it is important to
be precise enough in the measurements of the properties of the new scalar
particle in order to confirm its nature as the SM Higgs boson responsible for
electroweak symmetry breaking (EWSB). On the other hand, the Higgs boson mass
must be compatible with the requirements imposed by supersymmetry at the expense
of moving the SUSY scale above TeV energies. 
This relatively heavy Higgs boson
mass imposes indirect pressures on the supersymmetric spectrum. For instance,
there is a strong tension between LHC measurements and the need for a
fine-tuning that can be as large as 300 or more to accommodate a viable EWSB
mechanism in case of heavy higgsinos. It is nonetheless possible to get viable
scenarios with lighter higgsinos and a less extreme fine-tuning in some corners
of the parameter space~\cite{Cassel:2011tg,Baer:2014ica}.

One could assume that supersymmetry does not manifest itself as the minimal
supersymmetric standard model (MSSM), but feature instead an extended gauge
symmetry. This implies the presence of additional new particles that could alter
the exclusion limits derived in particular from measurements at the LHC in
proton-proton collisions at centre-of-mass energies $\sqrt{s}$ of 7, 8 and
13~TeV. Ideally, the new model would preserve all the attractive features of the
MSSM, resolve some of its outstanding issues, and allow for a parameter space
distinct for that of the MSSM in some regions.
One possible source of difference between an extended SUSY model and the MSSM
could be in the viable options for the LSP. In its minimal incarnation,
supersymmetry has one possible dark matter candidate, the neutralino which can
be an arbitrary admixture of binos, winos and higgsinos.

Dark matter searches can play an important role as probes for physics beyond the
SM, especially as providers of indirect information on the spectrum of the
models under investigation. We rely on these observations to investigate the
opportunities for natural DM candidates offered by extended supersymmetric
scenarios and to make use of dark matter data as a testing ground for extended
SUSY models. In one of the simplest extensions of the MSSM, the gauge group is
enlarged by an extra $U(1)^\prime$ symmetry.
This model minimally introduces a new gauge boson, a new singlet Higgs field,
and a right-handed neutrino, together with their superpartners. The right-handed
sneutrino can be the LSP and a viable DM candidate in particular thanks to its
interactions with the new gauge boson. This contrasts with the MSSM where
left-handed scalar neutrinos, which do not partake in strong and electromagnetic
interactions, cannot be possible candidates for DM as their interactions with
the $Z$ boson yield too high annihilation cross sections~\cite{Ellis:2010kf}.
In addition, the lightest neutralino, that can also be an acceptable DM
candidate, can exhibit novel properties due to its possible $U(1)^\prime$ bino
component. This would lead to additional annihilation channels which may imply
some dissimilarities with the MSSM neutralino LSP.

The possibility of adding an extra $U(1)^\prime $ gauge symmetry to the SM is
well-motivated in superstring constructions~\cite{Hewett:1988xc}, Grand Unified
Theories (GUTs)~\cite{Langacker:1998tc}, models of dynamical symmetry
breaking~\cite{Hill:2002ap}, little Higgs models~\cite{ArkaniHamed:2002qy,%
Han:2003wu}, and setups with large extra dimensions~\cite{Antoniadis:1990ew}.
Extra $U(1)^\prime$ groups generally arise from the breaking of an $SO(10)$ or
$E_6$ symmetry to the SM gauge symmetry. In supersymmetry, $U(1)^\prime$ models
also offer a solution to the MSSM fine-tuning issue that is mainly driven by the
bilinear $\mu$ term of the superpotential. This term is indeed simultaneously
responsible for the Higgs boson mass and for the higgsino masses. In the MSSM,
higgsinos are expected to be light, of ${\cal O}(100)$~GeV, while predictions
for a Higgs boson mass of about 125~GeV require supersymmetric masses of
${\cal O}(1)$~TeV or more. This raises questions about the nature of the $\mu$
parameter. $U(1)^\prime $ extensions of MSSM (UMSSM) suggest a solution to the
so-called $\mu $-problem by the introduction of an effective $\mu_{\rm eff}$
parameter dynamically generated by the vacuum expectation value (VEV) of a new
scalar field $S$ responsible for breaking the $U(1)^\prime$
symmetry~\cite{Cvetic:1997ky,Demir:2005ti}. While this resolution of the $\mu$
problem is similar to the one provided in the next-to-minimal supersymmetric
standard model (NMSSM)~\cite{Ellwanger:2009dp}, the $U(1)^\prime$ symmetry
additionally prevents from the appearance of cosmological domain
walls~\cite{Ellis:1986mq}.
Moreover, extra desirable features of UMSSM models are the absence of rapid
proton decay operators (of dimension four), the protection of all fields by
chirality and supersymmetry from acquiring high-scale masses, consistency with
anomaly cancellation, gauge-coupling unification, as well as family universality
that allow us to avoid flavour-changing neutral current
constraints~\cite{Langacker:2000ju}.

The aim of this article is to present a comprehensive study of all $U(1)^\prime$
models emerging from the breaking of an $E_6$ symmetry in contexts where
either a scalar neutrino or the lightest neutralino is the LSP. The former is
not a possibility available in the MSSM, and, as we shall see, not the most
natural solution in UMSSM models. There however exists a large variety of UMSSM
realisations where the lightest sneutrino, which contains a dominant
right-handed sneutrino component, is the LSP and where the observed dark matter
abundance can be explained while satisfying other experimental constraints. This
contrasts with left-handed sneutrino LSP scenarios which are excluded, as in the
MSSM, by a non-zero sneutrino hypercharge that leads to a too efficient DM
annihilation via a $Z$-boson exchange in the early Universe, and thus to a relic
abundance lower than the $\Omega_{\rm DM}h^2$ value measured by the
WMAP~\cite{Hinshaw:2012aka} and Planck~\cite{Ade:2013zuv} satellites. We explore
the UMSSM parameter space consistent with either a sneutrino or a neutralino
LSP, impose constraints from dark matter relic abundance and direct detection
experiments, and then investigate potential signals of the viable scenarios at
the LHC. We also address the compatibility of acceptable setups with
measurements of the anomalous magnetic moment of the muon $(g-2)_\mu$. The
differences between the two (sneutrino and neutralino LSP) scenarios are
outlined and we especially emphasise the challenges originating from the
fact that for most of the parameter space for which dark matter constraints are
satisfied, the expected LHC signals are not visible, while benchmark setups
yielding LHC signals that could be extracted from the SM background fail to
satisfy dark matter constraints.

Whereas previous phenomenological studies in specific UMSSM constructions
have appeared in Refs.~\cite{Ma:1986we,Ham:2007kc,Langacker:2008yv,
Belanger:2015cra,Belanger:2011rs,
Corcella:2014lha,Corcella:2012dw,BhupalDev:2012ru,Barger:2012ey,Chiang:2014yva,
Belanger:2017vpq}, our
analysis features new ingredients. It encompasses {\it all} possible $U(1)^\prime$
symmetries arising from the breaking of an $E_6$ symmetry, with the goal of
determining characteristic signals which discriminate them. We moreover include
all constraints arising from low-energy phenomena, updated results from the
$Z^\prime$ boson searches and from Higgs boson signal strength data. More
practically, we first perform a scan of the parameter space and then derive the
regions of the parameter space consistent with a viable sneutrino or
neutralino dark matter candidate. We then investigate the various signals that
could arise from dark matter experiments in order to pinpoint possible genuine
differences between the UMSSM realisations.

Our work is organised as follows. We review the properties of the supersymmetric
models featuring an extra $U(1)$ symmetry, or UMSSM models, in
Sec.~\ref{sec:model}. We then explore the corresponding parameter space and
determine the regions that exhibit a compatibility with the Higgs boson signal
strength and low-energy data in Sec.~\ref{sec:scan}, imposing the LSP to be
either a sneutrino or a neutralino. We next consider the associated
$Z^\prime$ boson phenomenology in Sec.~\ref{subsec:zprime} and the implications
for the anomalous magnetic moment of the muon in Sec.~\ref{subsec:muon_g-2}.
In Sec.~\ref{sec:sneutrinodm}, we focus on scenarios with a right sneutrino LSP
and analyse the dependence of the DM relic density on the $Z^\prime$ boson mass
as well as direct and indirect DM detection experiment signals. In
Sec.~\ref{sec:neutralinoLSP}, we investigate cases where the neutralino is the
LSP and again put an emphasis on the DM relic density, direct and
indirect detection constraints. We finally
discuss the prospects for observing UMSSM scenarios at
colliders in Sec.~\ref{subsec:benchmarks}. We summarise our findings and
conclude in Sec.~\ref{sec:conclusion}.

\section{UMSSM Models}
\label{sec:model}

\begin{figure}
  \includegraphics[scale=0.4]{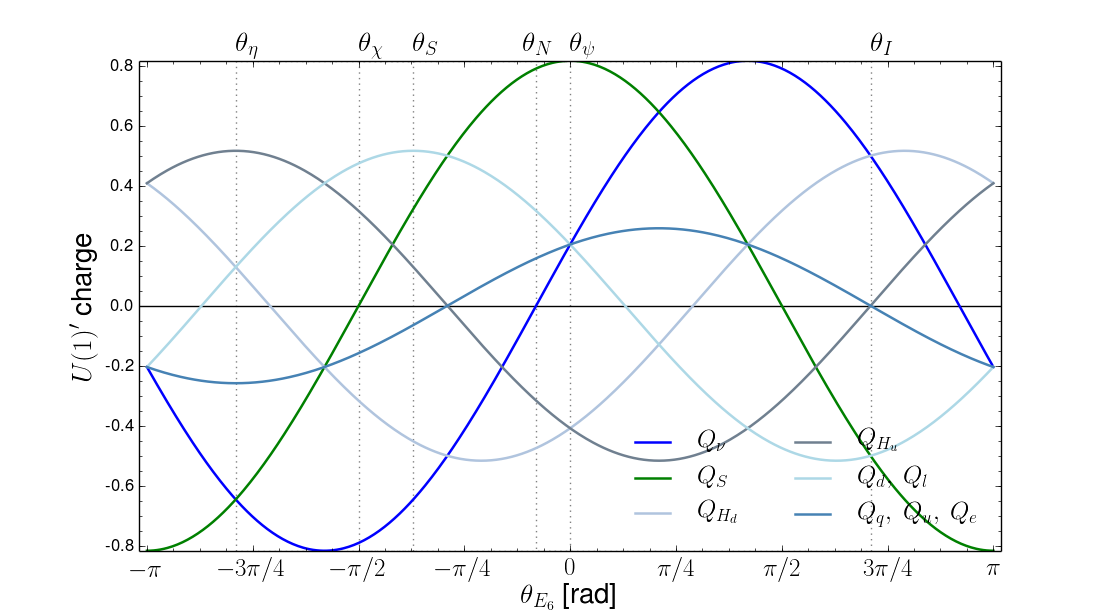}
  \caption{\label{fig:U1charge} \it Variation of the $U(1)^\prime$ charges of
    the various UMSSM superfields as a function of the $\theta_{E_6}$ mixing
    angle. The standard $U(1)'_\eta$, $U(1)'_\chi$, $U(1)'_S$, $U(1)'_N$,
    $U(1)'_\psi$ and $U(1)'_I$ models are identified by dotted
    vertical lines.}
\end{figure}

In this section, we briefly review the theoretical framework of minimal
$U(1)^\prime$-extended supersymmetric models that has been 
extensively discussed in Refs.~\cite{Erler:2000wu,Demir:2005ti,%
Langacker:2008yv}. The presence of the additional
gauge group introduces one extra neutral gauge boson $Z^\prime$ of mass $M_{Z'}$
together with
the corresponding gaugino superpartner $\lambda_{\tilde Z'}$. In their simplest
incarnations, UMSSM models also requires the presence of an additional
electroweak singlet superfield $S\equiv(s, \tilde s)$, charged under the
$U(1)^\prime$ symmetry, that is responsible for the breaking of the extended
symmetry group down to the electroweak group. The model field content moreover
includes two weak doublets of quark ($Q\equiv(q,\tilde q)$) and lepton ($L\equiv
(l,\tilde l)$) chiral supermultiplet as well as four weak singlets of up-type
quark ($U\equiv(u, \tilde u)$), down-type quark ($D\equiv(d,\tilde d)$), charged
lepton ($E\equiv(e,\tilde e)$) and right neutrino ($N\equiv(\nu_R,\tilde\nu_R)$)
chiral supermultiplets. The Higgs sector contains, in
addition to the $S$ singlet, two weak doublets of Higgs supermultiplets
($H_u\equiv(\tilde h_u, h_u)$ and $H_d\equiv(\tilde h_d, h_d)$), and the gauge
sector is similar to the one of the MSSM except for the $U(1)^\prime$
field. It thus includes a QCD ($G\equiv(g, \lambda_{\tilde G})$), weak
($W\equiv(w, \lambda_{\tilde W})$) and hypercharge
($B\equiv(b,\lambda_{\tilde B})$) gauge supermultiplets.

\begin{table}
  \renewcommand{\arraystretch}{1.6}
  \setlength\tabcolsep{5pt}
  \begin{tabular}{l||c|c|c|c|c|c}
  & $2\sqrt{10}Q^{'}_\chi $ & $ 2 \sqrt{6}Q^{'}_\psi $  & $2\sqrt{15}Q^{'}_\eta$
    & $ 2\sqrt{15}Q^{'}_S $ & $ 2Q^{'}_I $ & $ 2\sqrt{10}Q^{'}_N $ \\
  \hline \hline
  $\theta_{E_6} $ & $ -\pi/2$ & $ 0 $ & $ \arccos(\sqrt {5/8}) -\pi $ &
    $\arctan (\sqrt{15}/9) -\pi/2$ & $\arccos(\sqrt{5/8}) + \pi/2 $ &
    $\arctan\sqrt{15} -\pi/2$ \\
  \hline
  $ Q_{q,u,e} $ & -1     & 1  & -2 & -1/2 & 0  & 1 \\
  $ Q_{d,l} $ &  3     & 1  & 1  &   4  & -1 & 2 \\
  $ Q_\nu $ & -5   & 1  & -5 &  -5  & 1  & 0 \\
  $ Q_{H_u} $ & 2  & -2 & 4  &  1   & 0  & 2 \\
  $ Q_{H_d} $ & -2 & -2 & 1  & -7/2 & 1  & -3 \\
  $ Q_S $ & 0      & 4  & -5 &  5/2 & -1 & 5 \\
  \end{tabular}
  \caption{\it $U(1)^\prime$ charges of the UMSSM quark ($Q_q$, $Q_d$, $Q_u$),
    lepton ($Q_l$, $Q_e$, $Q_\nu$) and Higgs ($Q_{H_u}$, $Q_{H_d}$, $Q_S$)
    supermultiplets for the anomaly-free abelian group that could arise from the
    breaking of an $E_6$ symmetry. The value of the mixing angle $\theta_{E_6}
    \in [-\pi,\pi]$ is also indicated.\label{tab:u1charges}}
\end{table}

There are several possibilities for defining
the extra $U(1)^\prime$ symmetry. The most commonly used parameterisation
emerges from considering a linear combination of the maximal subgroups
$U(1)^\prime_\psi$ and $U(1)^\prime_\chi$ resulting from the breaking of a grand
unified $E_6$ gauge group~\cite{Deppisch:2007xu},
\be
  E_6 \longrightarrow SO(10) \otimes U(1)^\prime_\psi
   \longrightarrow \Big(SU(5) \otimes U(1)^\prime_\chi \Big)
     \otimes U(1)^\prime_\psi\ .
\ee
Introducing a mixing angle $\theta_{E_6}$, a general $U^\prime(1)$ charge
operator can be written from the respective $U(1)^\prime_\psi$ and
$U(1)^\prime_\chi$ charge operators $Q^\prime_\psi$ and $Q^\prime_\chi$ as
\be\label{eq:u1charge}
  Q^\prime(\theta_{E_6}) = Q^\prime_\psi\cos\theta_{E_6} -
   Q^\prime_\chi\sin\theta_{E_6}  \ .
\ee
In Fig.~\ref{fig:U1charge} we present the variation of the $U(1)^\prime$ charges
of the UMSSM quark, lepton and Higgs superfields as  functions of the mixing
angle $\theta_{E_6}$ that will be a key parameter of our analysis. We identify
by vertical lines the anomaly-free
$U(1)^\prime$ group choices denoted by $U(1)'_\eta$, $U(1)'_\chi$, $U(1)'_S$,
$U(1)'_N$, $U(1)'_\psi$ and $U(1)'_I$, and give the corresponding
charge and mixing angle values in Table~\ref{tab:u1charges}.

The UMSSM superpotential contains usual quarks and lepton Yukawa interactions
and reads, in the presence of a right-handed neutrino superfield $N$,
\be
  W = {\mathbf Y}_u\,U\,Q\,H_u\, - {\mathbf Y}_d \,D\,Q\,H_d\,-
     {\mathbf Y}_e \,E\,L\,H_d\,+  {\mathbf Y}_\nu\,L\,H_u\,N\,+
     \lambda\,H_u\,H_d\,S \ ,
\label{eq:superW}\ee
where the four Yukawa couplings ${\mathbf Y}_u$, ${\mathbf Y}_d$,
${\mathbf Y}_l$ and ${\mathbf Y}_\nu$ are $3\times 3$
matrices in flavour space and $\lambda$ represents the strength of the
electroweak Higgs singlet and doublet interactions. All indices are understood
but explicitly suppressed
for simplicity. After the breaking of the $U(1)^\prime$ symmetry, the scalar
component $s$ of the singlet superfield gets a VEV $v_S$ and the last
superpotential term of Eq.~\eqref{eq:superW} induces an effective $\mu$-term
with $\mu_{\rm eff}= \lambda v_S/\sqrt{2}$, allowing for the resolution of the
$\mu$-problem inherent to the MSSM~\cite{Kim:1983dt,Suematsu:1994qm,%
Cvetic:1996mf,Jain:1995cb,Demir:1998dm}. As in the MSSM, SUSY is softly broken
via the introduction of gaugino mass terms,
\be
  -{\cal L}_{\rm soft}^\lambda=\frac12 \Big(
    M_1 \lambda_{\tilde B} \cdot\lambda_{\tilde B} +
    M_2 \lambda_{\tilde W} \cdot\lambda_{\tilde W}+
    M_1' \lambda_{\tilde Z'}\cdot\lambda_{\tilde Z'}+
    M_3 \lambda_{\tilde g} \cdot\lambda_{\tilde g}+{\rm h.c.}\Big) \ ,
\ee
where the $M_i$ variables denote the various mass parameters, scalar mass
terms $m_i$,
\be
  -{\cal L}_{\rm soft}^\Phi = m_{H_d}^2h_d^\dag h_d + m_{H_u}^2 h_u^\dag h_u +
    m_{S}^2 s^2 + m^2_{\tilde Q}{\tilde q}^\dag\tilde q +
    m^2_{\tilde d}{\tilde d}^\dag{\tilde d} +
    m^2_{\tilde u}{\tilde u}^\dag\tilde u +
    m^2_{\tilde L}{\tilde l}^\dag \tilde l + 
    m^2_{\tilde e}{\tilde e}^\dag{\tilde e}+
    m^2_{\tilde \nu}{\tilde \nu}_R^\dag{\tilde \nu}_R \ ,
\ee
and trilinear interactions
featuring a structure deduced from the one of the superpotential,
\be\bsp
  -{\cal L}_{\rm soft}^W =&\ A_\lambda\, s\, h_u\, h_d\, -
    A_d\, {\tilde d}^\dag\, {\tilde q}\, h_d\, -
    A_e\, {\tilde e}^\dag\, {\tilde l}\, h_d\, +
    A_u\, {\tilde u}^\dag\, {\tilde d}\, h_u\, +{\rm h.c.}\ ,
\esp\ee
where the $A_i$ parameters stand for the soft couplings.

After the breaking of the UMSSM gauge symmetry down to electromagnetism, all
neutral components of the scalar Higgs fields get VEVs,
$\langle h^0_u\rangle = v_u/\sqrt{2}$, $\langle h^0_d\rangle = v_d/\sqrt{2}$ and
$\langle s\rangle = v_S/\sqrt{2}$. As a consequence, UMSSM models can easily
lead to neutrino masses that are consistent with neutrino oscillation data
through an implementation of a see-saw mechanism~\cite{Minkowski:1977sc,%
Mohapatra:1979ia,Schechter:1980gr,Schechter:1981cv}. The exact details depend on
the form of the extra
$U(1)^\prime$ symmetry~\cite{Kang:2004ix}, and viable models can be constructed
to contain
Dirac-type~\cite{Demir:2006jj} or Majorana neutrino masses~\cite{Demir:2007dt}.
The symmetry breaking mechanism additionally induces the mixing of fields
carrying the same spin, colour and electric charge quantum numbers, and
the gauge eigenbasis has to be rotated to the physical basis. Contrary to the
MSSM where the tree-level SM-like Higgs-boson mass is bound by the $Z$-boson
mass $M_Z$ so that large stop masses and/or trilinear $A_t$ couplings are
required for pushing the loop corrections to a large enough
value~\cite{Heinemeyer:2011aa}, the singlet field provides new tree-level
$F$-term contributions that naturally stabilise the SM-like Higgs boson mass
$M_h$ to a greater value more easily in agreement with the measured experimental
value of 125~GeV~\cite{Ross:2012nr}. For any further details on the resulting
particle
spectrum, we refer to Refs.~\cite{Cvetic:1997ky,Demir:2005ti,Belanger:2015cra}.

The UMSSM Lagrangian introduced above exhibit numerous parameters, in
particular within its soft SUSY-breaking part. To reduce the
dimensionality of the parameter space, we assume that the SUSY-breaking
mechanism originates from minimal supergravity so that unification relations
amongst the soft masses can be imposed at the GUT scale where an $E_6$ gauge
symmetry is realised. We however deviate from the most minimal model by
maintaing the freedom to choose the details of the lepton and neutrino sector,
which guarantees that a sneutrino could be the LSP.
More details are given in the following section.

\section{Parameter Space Scan and Constraints}
\label{sec:scan}

\subsection{Technical setup}
\label{sec:constr}

We perform a scan of the UMSSM parameter space in order to determine regions in
which either a sneutrino or a neutralino is the LSP and thus a potential
dark matter candidate. We focus on the six anomaly-free UMSSM realisations
introduced in the previous section. More precisely, we generate the particle
spectrum by making use of \textsc{Sarah} version 4.6.0~\cite{Staub:2013tta} and
\textsc{SPheno} version 3.3.8~\cite{Porod:2011nf}. Predictions for the dark
matter observables are then achieved with \textsc{micrOMEGAs} version
4.3.1~\cite{Belanger:2014vza}, and the properties of the Higgs sector are
evaluated with \textsc{HiggsBounds} version 4.3.1~\cite{Bechtle:2008jh} and
\textsc{HiggsSignals} version 1.4.0~\cite{Bechtle:2013xfa}. The interfacing of
the various programmes and our numerical analysis have been implemented within
the \textsc{pySLHA} package, version 3.1.1~\cite{Buckley:2013jua}.

We make use of GUT-inspired relations to simplify the size of the
parameter space. The considered set of free parameters is given by
\be
  M_{1/2},\quad  M_0,\quad m_{\tilde e},\quad
  m_{\tilde L}, \quad m_{\tilde \nu},\quad
  \tan\beta=\frac{v_u}{v_d}, \quad \mu_{\rm eff},\quad
  A_0, \quad A_\lambda,\quad Y_\nu,\quad
  M_{Z^\prime} \quad\text{and}\quad \theta_{E_6} \ ,
\label{eq:prms}\ee
where we have enforced a unification relation at the GUT scale relating the
$U(1)^\prime$, hypercharge, weak and QCD gaugino soft masses $M_1^\prime=M_1=M_2
=M_3=M_{1/2}$ as well as the hypercharge, weak and $U(1)^\prime$ gauge couplings
$g_1 = g_2 = g^\prime\sqrt{3/5}$. Constraining the SUSY scale to be below 5~TeV,
renormalization group evolution implies, at the SUSY scale, that $6M_1 \approx
3M_2 \approx M_3$. We have moreover required that all squark soft masses and
trilinear couplings respectively unify to common values $M_0$ and $A_0 Y_q$ at
the GUT scale, the slepton and sneutrino masses $m_{\tilde e}$, $m_{\tilde L}$
and $m_{\tilde \nu}$ being kept independent whereas the leptonic trilinear
coupling $A_e$ is taken vanishing. The neutrino Yukawa matrix is finally fixed
to a diagonal matrix with entries equal to $10^{-11}$.

\begin{table}
  \setlength\tabcolsep{7pt}
  \renewcommand{\arraystretch}{1.4}
  \begin{tabular}{c|c||c|c}
    Parameter      & Scanned range& Parameter      & Scanned range\\
    \hline
    $M_0$          & $[0, 3]$~TeV & $\mu$          & $[-2, 2]$~TeV\\
    $M_{1/2}$      & $[0, 5]$~TeV & $A_\lambda$    & $[-7, 7]$~TeV\\
    $A_0$          & $[-3, 3]$~TeV& $M_{Z'}$       & $[1.98, 5.2]$~TeV\\
    $\tan\beta$    & $[0, 60]$    & $m^2_{\tilde\nu}$& $[-6.8, 9]$~TeV$^2$\\
    $\theta_{E_6}$ & $[-\pi, \pi]$&
      $m^2_{\tilde e, \tilde l}$ & $[0, 1]$~TeV$^2$\\
  \end{tabular}
  \caption{\label{tab:scan_lim} \it Ranges over which we allow the free
    parameters of Eq.~\eqref{eq:prms} to vary.}
\end{table}

\begin{table}
  \setlength\tabcolsep{7pt}
  \renewcommand{\arraystretch}{1.6}
  \begin{tabular}{l|c|c||l|c|c}
    Observable & Constraints & Ref. & Observable & Constraints & Ref.\\
    \hline
    $M_h$ & $ 125.09 \pm 3 $ GeV (theo) & \cite{Chatrchyan:2012xdj} &
      $\chi^2(\hat{\mu})$ & $\leq 70 $ & - \\
    $|\alpha_{ZZ^\prime}| $& $ O(10^{-3}) $ & \cite{Erler:2009jh} &
      $M_{\tilde{g}} $& $ > 1.75 $ TeV & \cite{Khachatryan:2016xvy}\\
    $M_{\chi^0_2}$ & $ > 62.4$ GeV & \cite{Olive:2016xmw} & 
      $M_{\chi^0_3} $ & $ > 99.9 $ GeV & \cite{Olive:2016xmw}\\
    $M_{\chi^0_4} $      & $ > 116 $ GeV & \cite{Olive:2016xmw} &
      $M_{\chi^\pm_i} $    & $ > 103.5 $ GeV & \cite{Olive:2016xmw}\\
    $M_{\tilde{\tau}} $ & $ > 81 $ GeV & \cite{Olive:2016xmw} &
      $M_{\tilde{e}} $ & $ > 107 $ GeV & \cite{Olive:2016xmw} \\
    $M_{\tilde{\mu}} $& $ > 94 $ GeV & \cite{Olive:2016xmw} &
      $M_{\tilde{t}} $& $ > 900 $ GeV & \cite{Aaboud:2016lwz}\\
    BR$(B^0_s \to \mu^+\mu^-) $ & $[1.1\times10^{-9},6.4\times10^{-9}]$ &
       \cite{Aaij:2012nna} & $\displaystyle  \frac{{\rm BR}(B \to \tau\nu_\tau)}
       {{\rm BR}_{SM}(B \to \tau\nu_\tau)} $ &$  [0.15,2.41] $ &
       \cite{Asner:2010qj}\\
    BR$(B^0 \to X_s \gamma) $ &$  [2.99,3.87]\times10^{-4} $ &
       \cite{Amhis:2016xyh}\\
  \end{tabular}
  \caption{\label{tab:constraints} \it Experimental constraints imposed within
  our scanning procedure in order to determine the parameter space regions of
  interest.}
\end{table}

Our parameter space investigation relies on the Metropolis-Hasting sampling
method where the free parameters of Eq.~\eqref{eq:prms} are allowed to vary in
the ranges given in Table~\ref{tab:scan_lim}, the lower bound on the mass of the
$Z'$ boson being the minimum value allowed for any choice of the $U(1)^\prime$
symmetry (and corresponds to the $ U(1)^\prime_{\eta}$ case). 
 This mass has been taken smaller than the one quoted in the 2016
Particle Data Group review~\cite{Olive:2016xmw} in order to allow for
significant branching fractions for the $Z'$ boson decays into a pair of
supersymmetric particles \cite{Corcella:2013cma}. We have
retained scenarios for which the predictions for the observables listed in
Table~\ref{tab:constraints} agree with the experimental data. Constraints
arising from the Higgs sector, namely a theory-experiment agreement for the
Higgs boson mass, the gluon and vector boson fusion Higgs boson production
cross-sections, and the Higgs signal strengths, have been applied by using
\textsc{HiggsBounds} and \textsc{HiggsSignals}. This is achieved by evaluating
the Higgs boson production rate in the gluon and vector boson fusion channels
with the \textsc{SusHi} program version 1.5~\cite{Harlander:2012pb} and by then
comparing the predictions to $\sigma(gg\rightarrow h)= 19.27^{+1.76}_{-4.44}$~pb
and $\sigma(VV\rightarrow h)= 1.55^{+0.058}_{-0.039}$~pb for a centre-of-mass
energy of 8 TeV and $\sigma(gg\rightarrow h) = 50.74^{+4.68}_{-11.6}$~pb for a
centre-of-mass energy of 14~TeV~\cite{Heinemeyer:2013tqa}. We next derive a
$\chi^2(\hat\mu)$ quantity for each, estimating the deviation from the
experimental data, the sum of which we enforce to be
smaller than 70. We have moreover severely restricted any possible kinetic
mixing between the $Z$ and the $Z^\prime$ bosons, and required that the
associated mixing angle $\alpha_{ZZ^\prime}$ is of the order of $10^{-3}$.
We have additionally verified that predictions for the gluino mass $M_{\tilde g}$, the neutralino and chargino masses $M_{\tilde\chi^0_i}$ and $M_{\tilde\chi^\pm_i}$, the slepton masses $M_{\tilde e}$, $M_{\tilde\mu}$ and $M_{\tilde\tau}$ and the stop mass $M_{\tilde t}$ satisfy the experimental bounds~\cite{Olive:2016xmw}. We have also imposed constraints arising from $B$-physics that are related to rare $B$-meson decays, and checked that the three branching ratios BR$(B^0_s \to \mu^+\mu^-)$, BR$(B \to \tau\nu_\tau)$ and
BR$(B^0 \to X_s \gamma)$ agree with existing data.

\subsection{General considerations and phenomenology of the Higgs sector}
\begin{figure}
  \centering
  \includegraphics[scale=0.42]{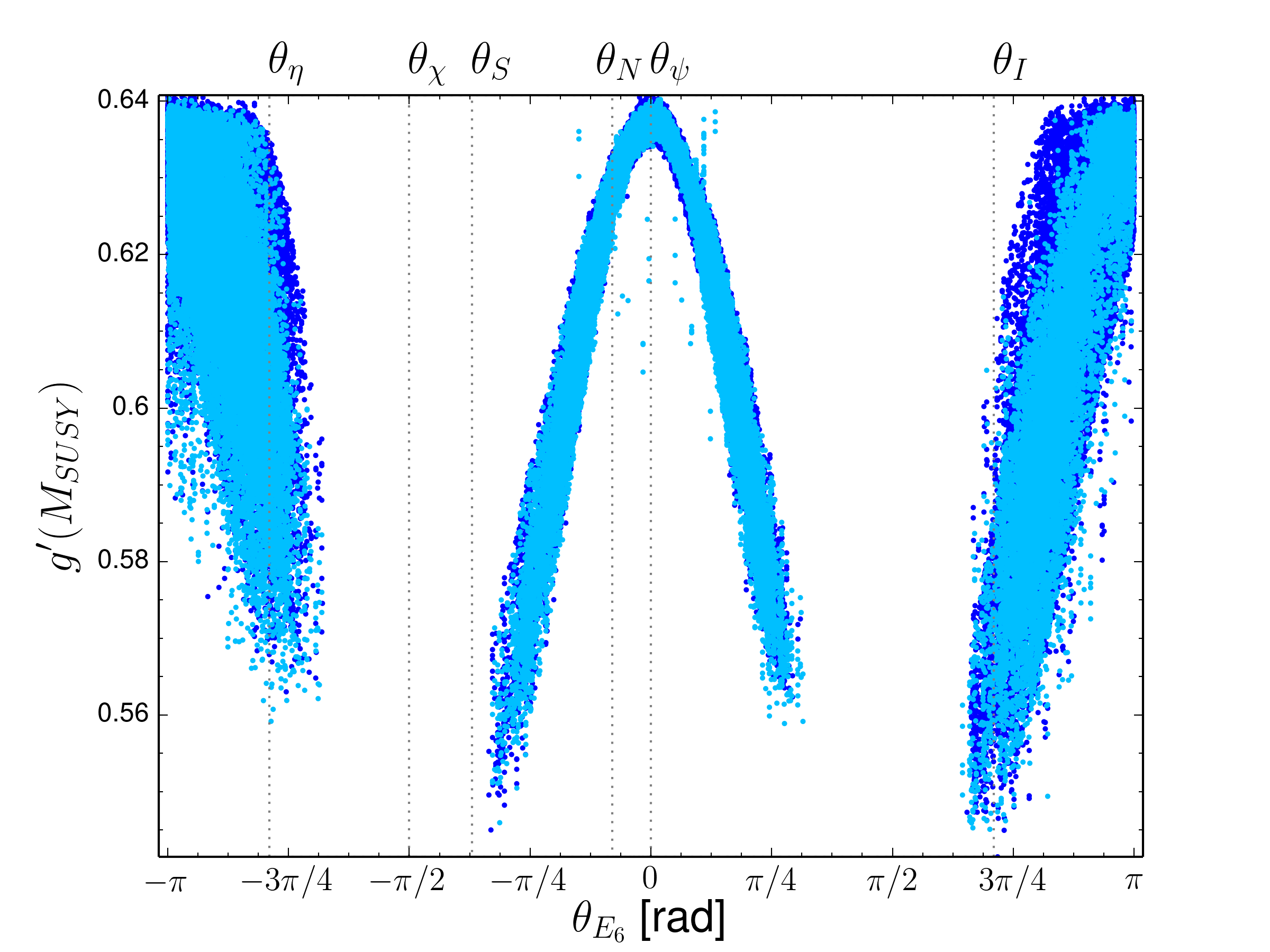}\hspace{.5cm}
  \includegraphics[scale=0.42]{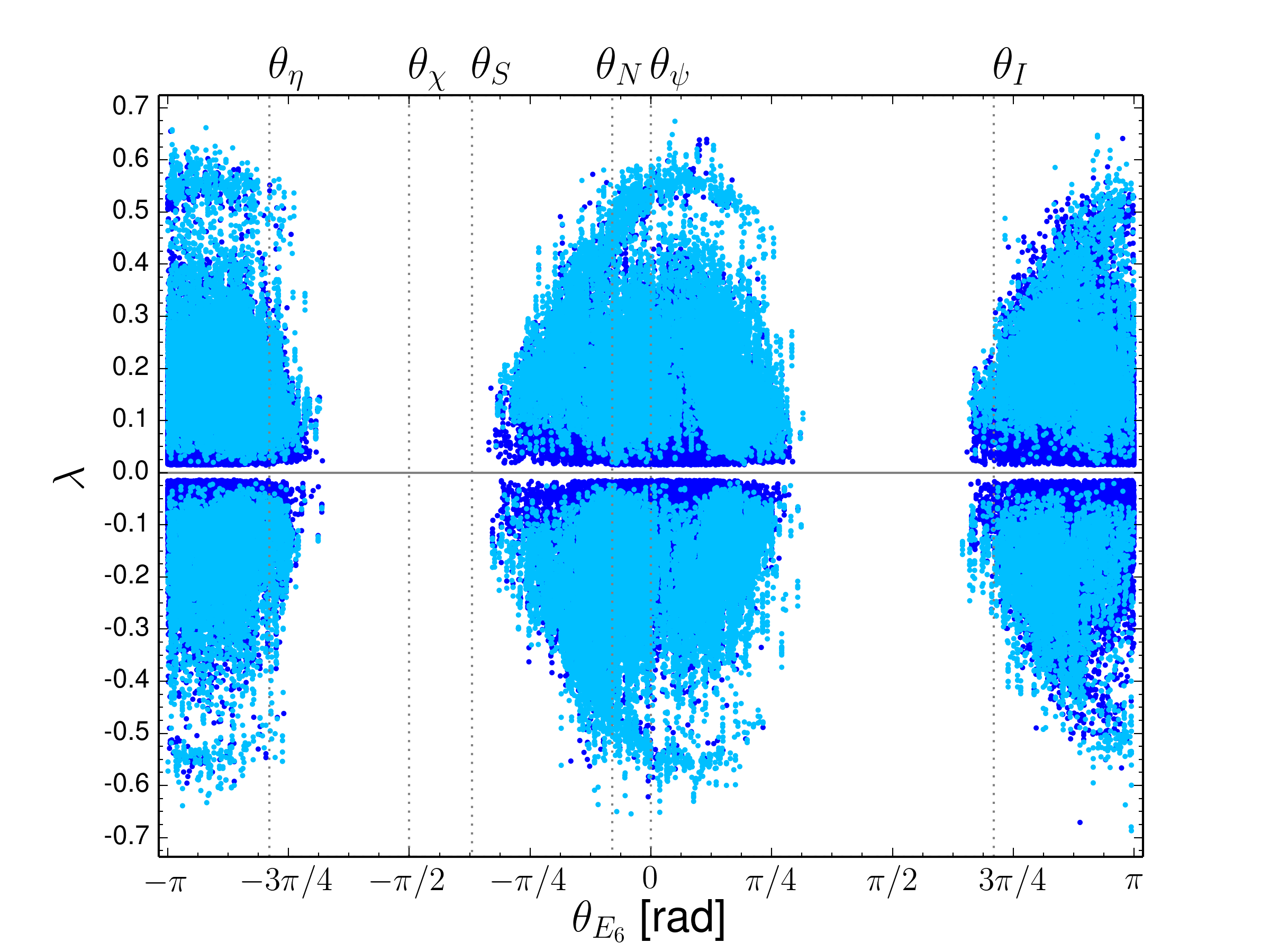}\\
  \includegraphics[scale=0.42]{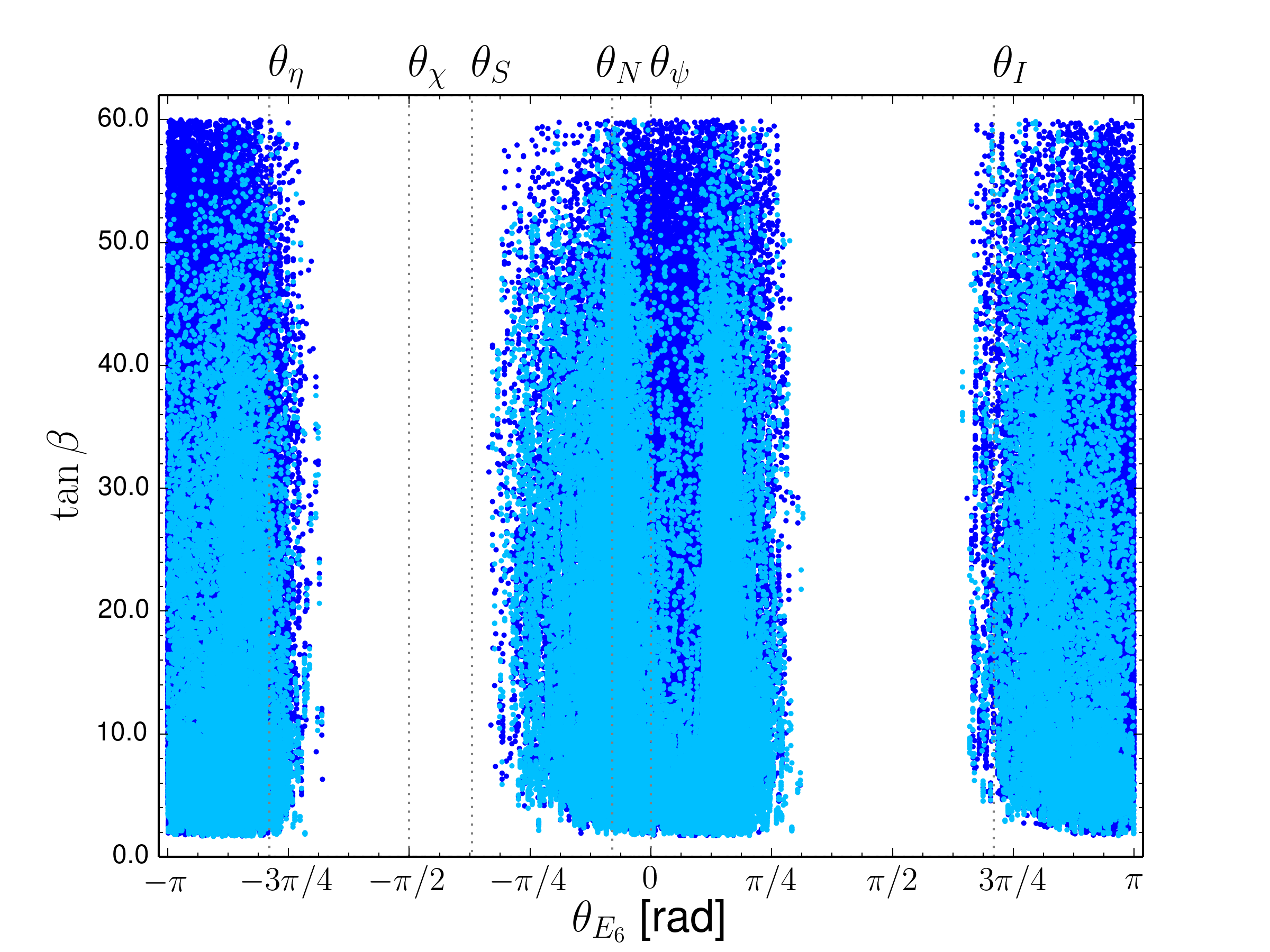}\hspace{.5cm}
  \includegraphics[scale=0.42]{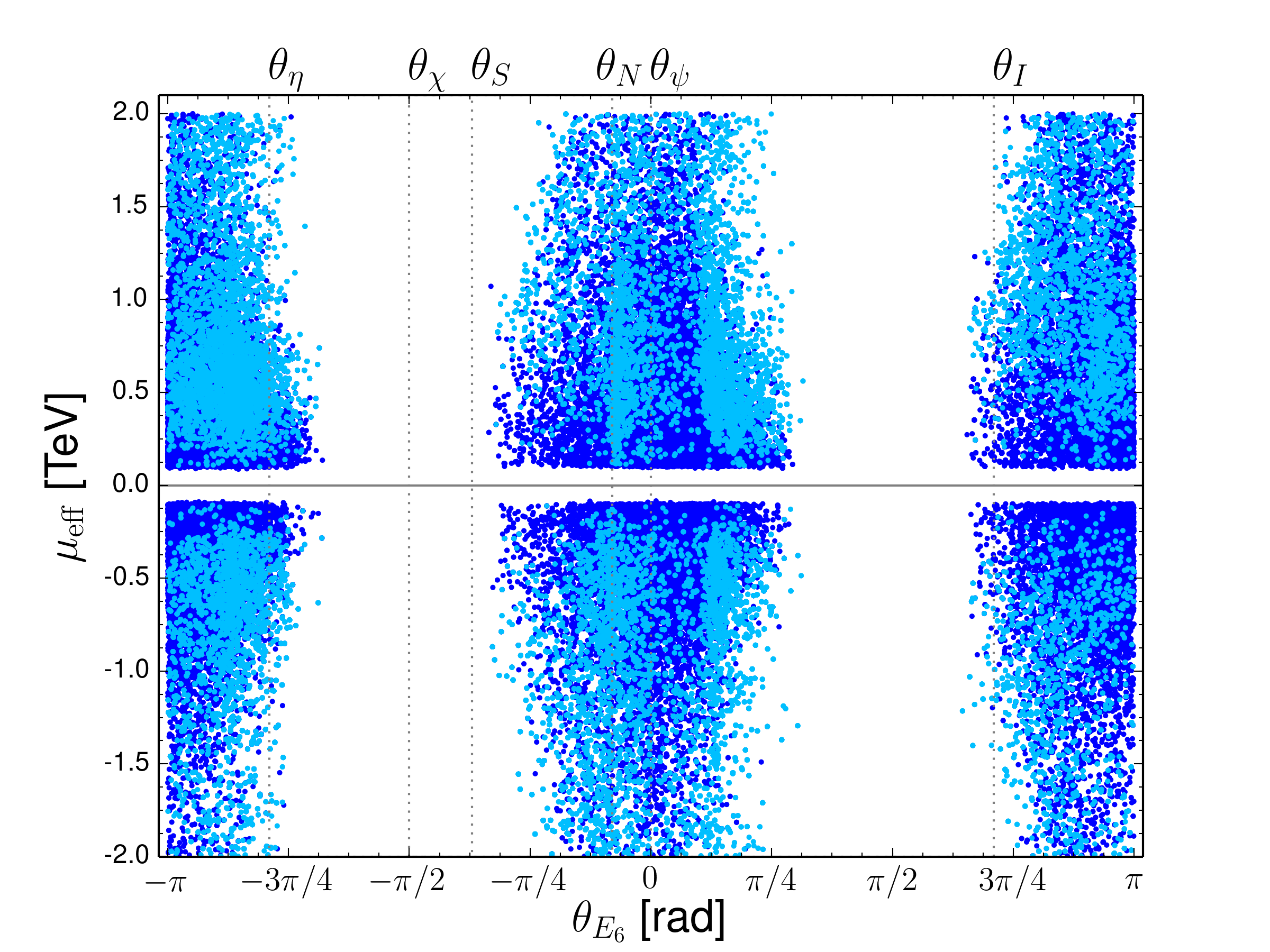}
  \caption{\it Distributions in the UMSSM parameter space of the scenarios in
    agreement with the constraints imposed on Sec.~\ref{sec:constr}. Results are
    projected into the $(\theta_{E_6}, g')$ (upper left panel), $(\theta_{E_6},
    \lambda)$ (upper right panel),  $(\theta_{E_6}, \tan\beta)$ (lower left
    panel) and  $(\theta_{E_6}, \mu_{\rm eff})$ planes. The light and dark blue
    points respectively represent scenarios in which the lightest sneutrino and
    the lightest neutralino is the LSP.}
  \label{fig:MCMC1}
\end{figure}

In Fig.~\ref{fig:MCMC1} we present the results of our scan. We project the
ensemble of accepted scenarios onto four two-dimensional planes in order to
exhibit possible correlations between the $U(1)^\prime$ mixing angle and the
$U(1)^\prime$ coupling $g'$ (upper left panel), the superpotential parameter
$\lambda$ (upper right panel), $\tan\beta$ (lower left panel) and the effective
$\mu_{\rm eff}$ parameter (lower right panel). We moreover distinguish the
classes of scenarios for which the LSP is a sneutrino (light blue points)
and a neutralino (dark blue points).

In the upper left panel of Fig.~\ref{fig:MCMC1}, we observe that the $g'$
coupling is in general large, which indicates that the $U(1)^\prime$
interactions must be strong to satisfy all the imposed constraints. Whereas the
value of $g^\prime$ is maximal in the context of $U(1)^\prime_\psi$ models, it
is generally highly dependent on many other
parameters so that a large range of values can be probed, regardless of the
precise choice of $\theta_{E_6}$. We however observe that $\theta_{E_6}$
values around $\pm\pi/2$ do not offer any option for a phenomenologically viable
scenario. This in particular disfavours the $U(1)^\prime_S$ and
$U(1)^\prime_\chi$ models, as already suggested by the results of
Fig.~\ref{fig:U1charge} where the $U(1)^\prime$ charge of the electroweak
singlet approaches zero for $ \theta_{E_6}\approx\pm\pi/2$. In this case, the
scalar field $s$ is not sufficient to break the $U(1)^\prime$ symmetry and one
cannot construct any predictable scenario.

The general features of the Higgs sector are then analysed in the three other
panels of Fig.~\ref{fig:MCMC1}. The distribution of the $\lambda$ parameter as a
function of the $\theta_{E_6}$ angle depicts how the weak singlet and doublets
of Higgs fields mix. This information is also represented in the lower right
panel of the figure where the $\lambda$ parameter is traded for the effective
$\mu_{\rm eff}$ parameter to which it is proportional. While all possible values
(different enough from zero and below 0.6 in absolute value) are in principle
possible regardless of the mixing angle value, the anomaly-free $U(1)^\prime_I$
model has the particularity to forbid $|\lambda| \gtrsim 0.3$. This stems from
the structure of the $U(1)^\prime$ charges that are small or vanishing for
several
supermultiplets and the lower bound on the $Z'$ mass in the scanning procedure
that both forbid $\lambda$ to be too large. A similar effect being also
observed for $\theta_{E_6}\approx -\pi/4$. The $\lambda$ parameter must
additionally be sufficiently large, in absolute value, to induce a successful
EWSB so that $\lambda$ values close to zero are forbidden.

While in general a sneutrino LSP can be obtained for any value of $\tan\beta$,
this turns out to be easier in the case of $U(1)^\prime_N$ models. These are
scenarios where the right neutrino supermultiplet is not charged under the
extended gauge symmetry, and right sneutrino masses do not therefore receive
any contribution from the $D$-terms and mostly arise from the independent soft
mass terms. As a result, one gets more freedom on $\tan\beta$ that can be
consequently lower. A similar feature, but less pronounced, can be observed for
other $\theta_{E_6}$ values where a combination of several zero $U(1)^\prime$
charges leads to the same conclusions.

\begin{figure}
  \includegraphics[scale=0.42]{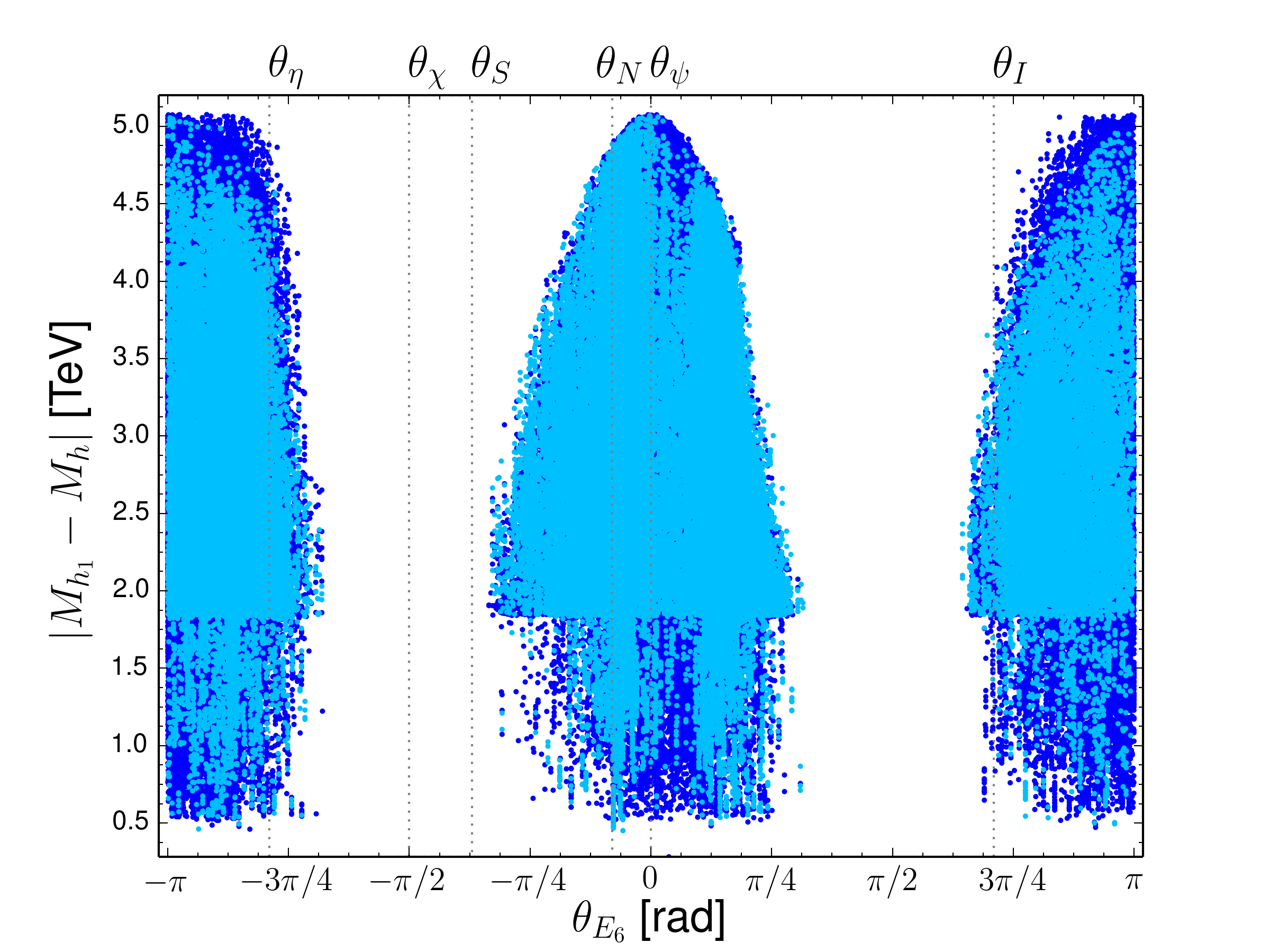}\hspace{0.5cm}
  \includegraphics[scale=0.42]{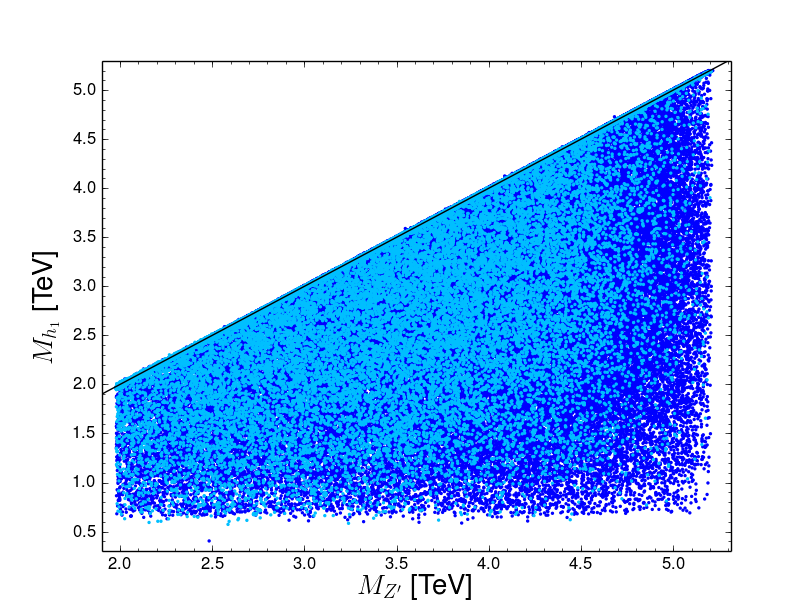}
  \caption{\it Same as in Figure~\ref{fig:MCMC1} but for projection in the
  $(\theta_{E_6}, |M_{h_1}-M_h|)$ (left panel) and $(\theta_{E_6}, M_{Z'})$
  (right panel) planes.}
  \label{fig:MCMC3}
\end{figure}

We further investigate the properties of the Higgs sector in
Fig.~\ref{fig:MCMC3}
where we present both the mass difference between the SM-like Higgs boson and
the next-to-lightest Higgs boson, $|M_{h_1} - M_h|$, in the left panel of the
figure and the dependence of $M_{h_1}$ on the $Z'$-boson mass in the right panel
of the figure. As the singlet VEV drives the $Z'$ boson mass, the second
lightest Higgs boson has a mass of at most roughly the $Z'$-boson mass and is in
this case singlet-dominated. In the lighter cases, it is mostly a doublet
admixture and thus MSSM-like. There are a few scenarios featuring a
sneutrino LSP where the second Higgs and the $Z'$ bosons are almost degenerate,
but any hierarchy can however be realised. The second Higgs boson is however at
least 500~GeV heavier than the SM-like Higgs boson, which originates from the
Higgs mixing pattern and the minimum value of the singlet VEV $v_S$ (that stems
from the $M_{Z'}$ lower limit imposed in our scan). Once again, smaller
$\lambda$ values obtained for  the case of the $U(1)^\prime_I$ scenario impact the
spectrum and $M_{h_1}$ is in general consequently smaller, the effects driven by
the large $v_S$ value being tamed by the smaller $\lambda$ value.

\subsection{$Z^\prime$ phenomenology}
\label{subsec:zprime}

\begin{figure}
  \centering
  \includegraphics[scale=0.42]{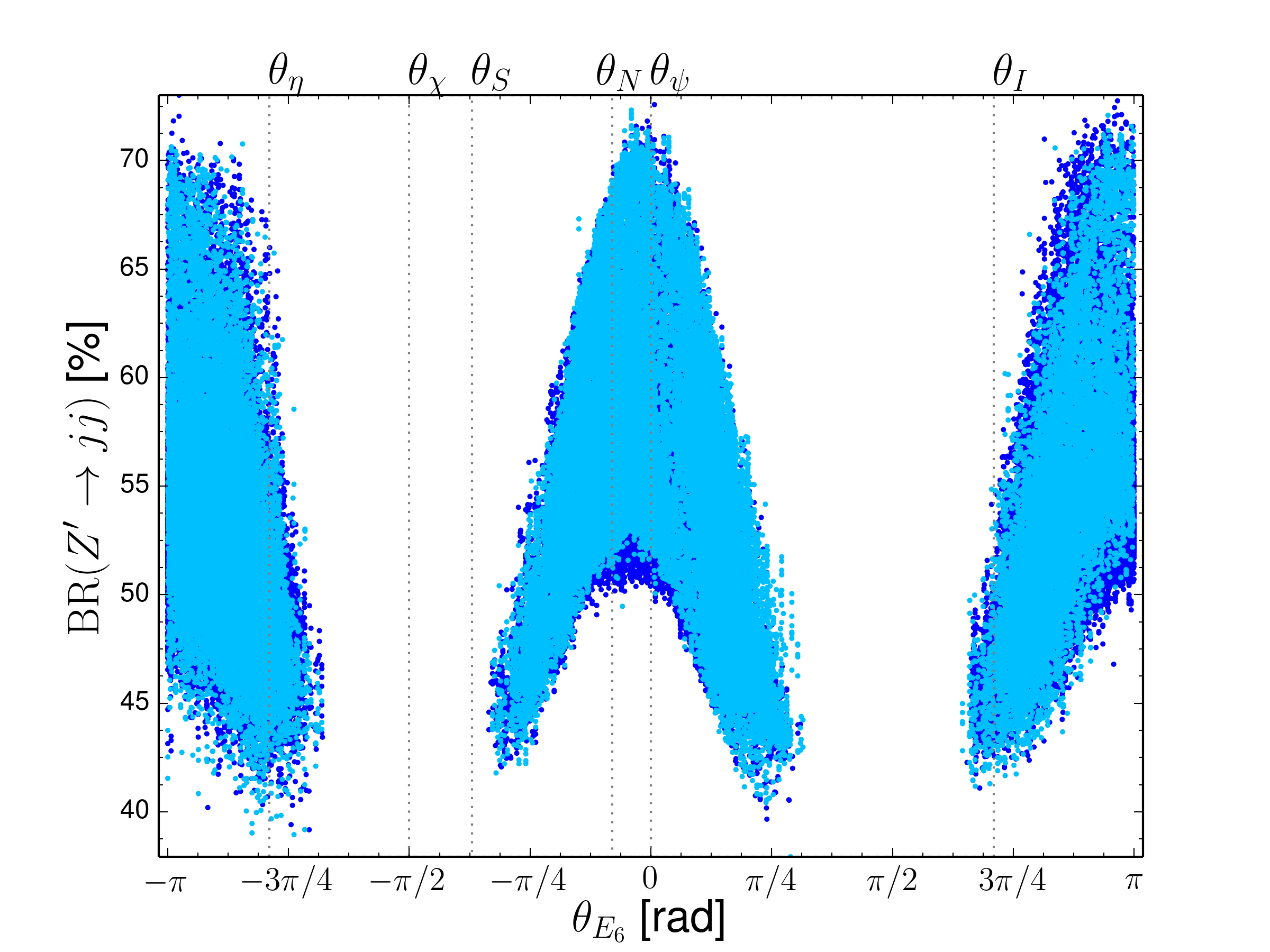}\hspace{0.5cm}
  \includegraphics[scale=0.42]{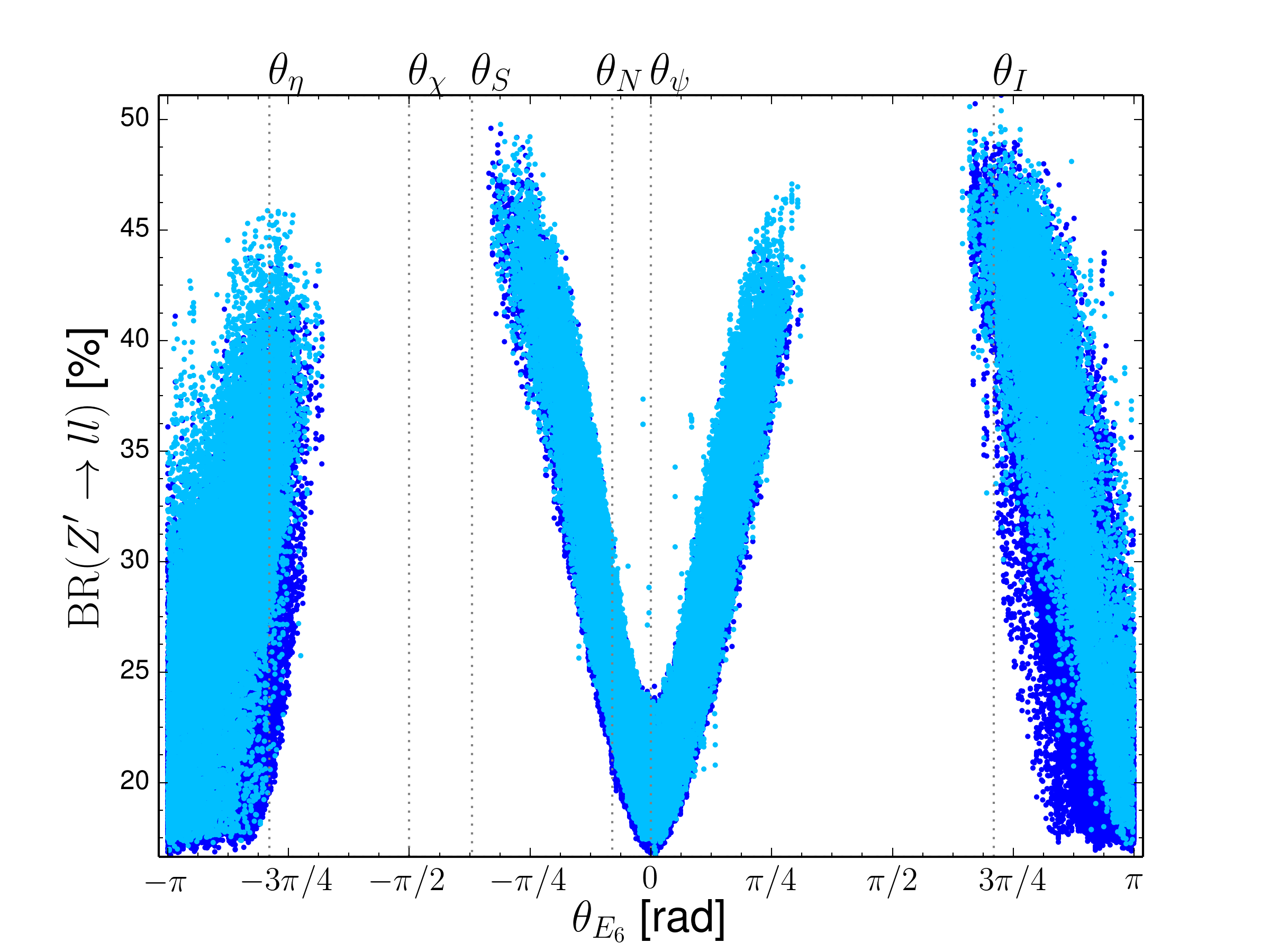}\\
  \includegraphics[scale=0.42]{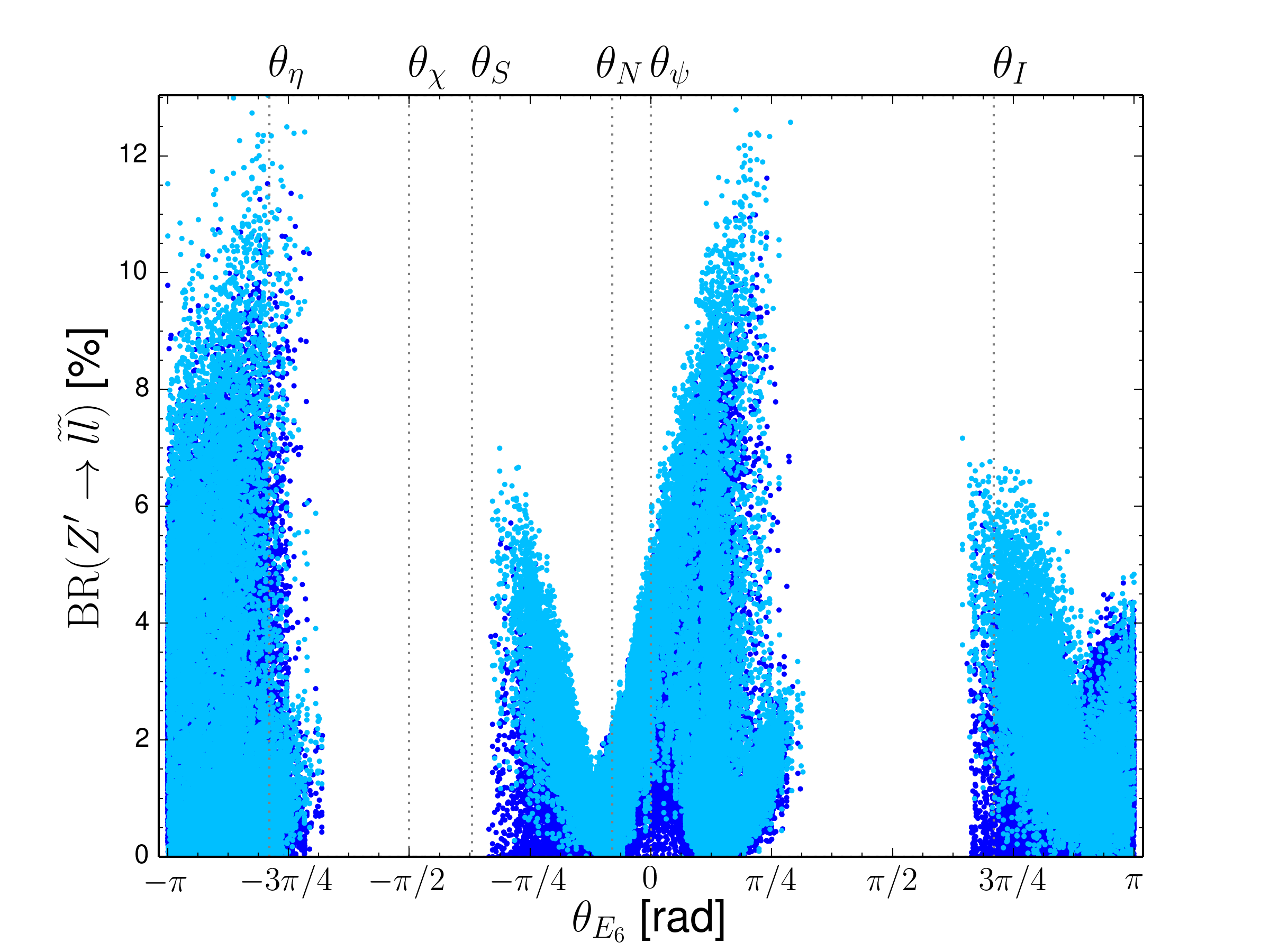}\hspace{0.5cm}
  \includegraphics[scale=0.42]{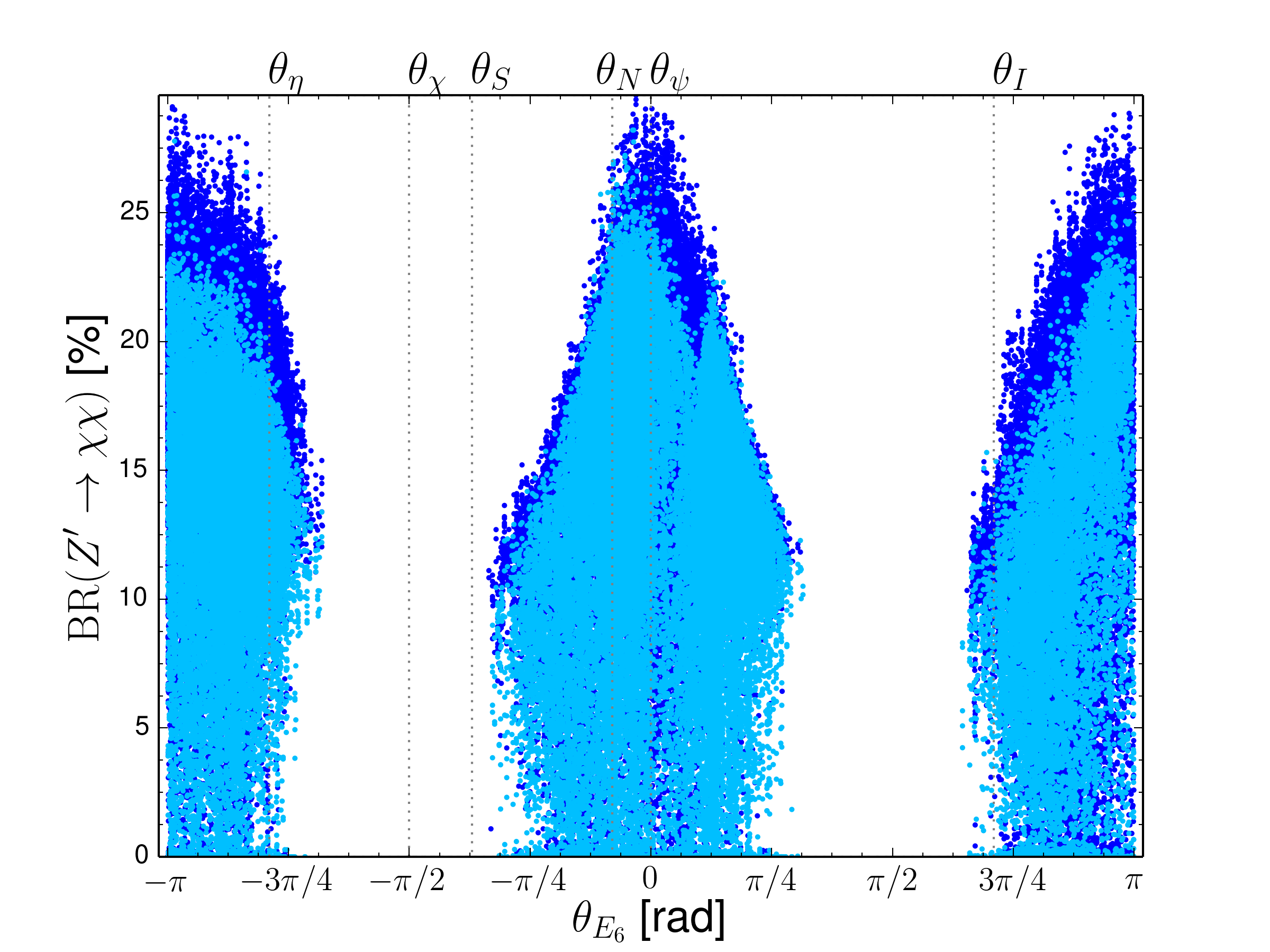}
  \caption{\it Same as in Fig.~\ref{fig:MCMC1} but for the branching ratios of
    the $Z'$ boson for several decay channels, namely the $Z'$ decays into a
    pair of jets (upper left), a pair of leptons (upper right), a pair of
    sleptons (lower left) and a pair of neutralinos or charginos (lower right).}
  \label{fig:ZpBR}
\end{figure}

Typical $Z'$ phenomenology can be dramatically different in the presence of
supersymmetry, in particular due to the existence of new $Z'$ decay channels
into pairs of SUSY particles. This is illustrated in Fig.~\ref{fig:ZpBR} where
we analyse different options for the $Z'$ decays as a function of the mixing
angle $\theta_{E_6}$.

Our results show that there is very little hope to be able to use $Z'$ decay
rates to differentiate $U(1)^\prime$ models. Decays into slepton pairs are
consistently small, while leptonic channels, that are also present in
non-supersymmetric cases, exhibit branching ratios ranging from 0 to about 50\%.
A leptophobic behaviour emerges for specific mixing angles, but these
features can be reproduced for other realisations where a large leptonic $Z'$
branching fraction is as well common. This nevertheless leads to one of the
most promising channels to look for a sign of $U(1)'$ new physics, by bump
hunting in the dilepton mass distribution for LHC events featuring two
opposite-sign final state leptons, provided the branching is large enough.
The same conclusion holds for the dijet decay mode that corresponds to the
preferred $Z'$ decay mode, regardless of the value of $\theta_{E_6}$. The only
limiting factor is, both for the dilepton and dijet case, the $Z'$ mass driving
the production cross section and the associated phase space suppression in the
heavy case.

In the lower right panel of the figure, we investigate the magnitude of the $Z'$
branching fraction into a pair of neutralinos or charginos. Such decays can
often be abundant, with a branching ratio reaching about 20\%, and yield a $Z'$
signature made of both leptons and missing energy. This potentially allows for
the distinction of SUSY and non-SUSY $Z'$-bosons.

\subsection{The anomalous magnetic moment of the muon}
\label{subsec:muon_g-2}
Pioneering results from the BNL E821 experiment~\cite{Bennett:2006fi}, their
improvements at the FNAL E989 experiment~\cite{Grange:2015fou} and the
anticipated results obtained from the J-PARC E34 experiment~\cite{Saito:2012zz}
have provided a very precise measurement of the anomalous magnetic moment of the
muon $(g-2)_\mu$. The measured value departs by about $3\sigma$ from the SM
expectation,
\be 
  a_\mu^{\rm SM}=116591828(2)(43)(26) \times 10^{-11} \ ,
\ee
which constitutes a challenge for beyond the SM model building. In the UMSSM
framework, both the presence of the extra gauge boson and a neutral and charged
(s)lepton sector in the presence of a sneutrino or neutralino LSP can have a
drastic impact on the anomalous magnetic moment of the muon via loop-induced
contributions. As the LSP is often much lighter than the $Z'$ boson, the
corresponding SUSY contributions are expected to be more important than any
additional $Z^\prime$ contribution. As in the MSSM, new physics effects on
$(g-2)_\mu$ are therefore mostly depending on $\tan\beta$ and the effective
$\mu_{\rm eff}$ parameter, which determine the higgsino masses and the
fermion and sfermion interactions with the higgs(ino) sector.

\begin{figure}
  \centering
  \includegraphics[scale=0.42]{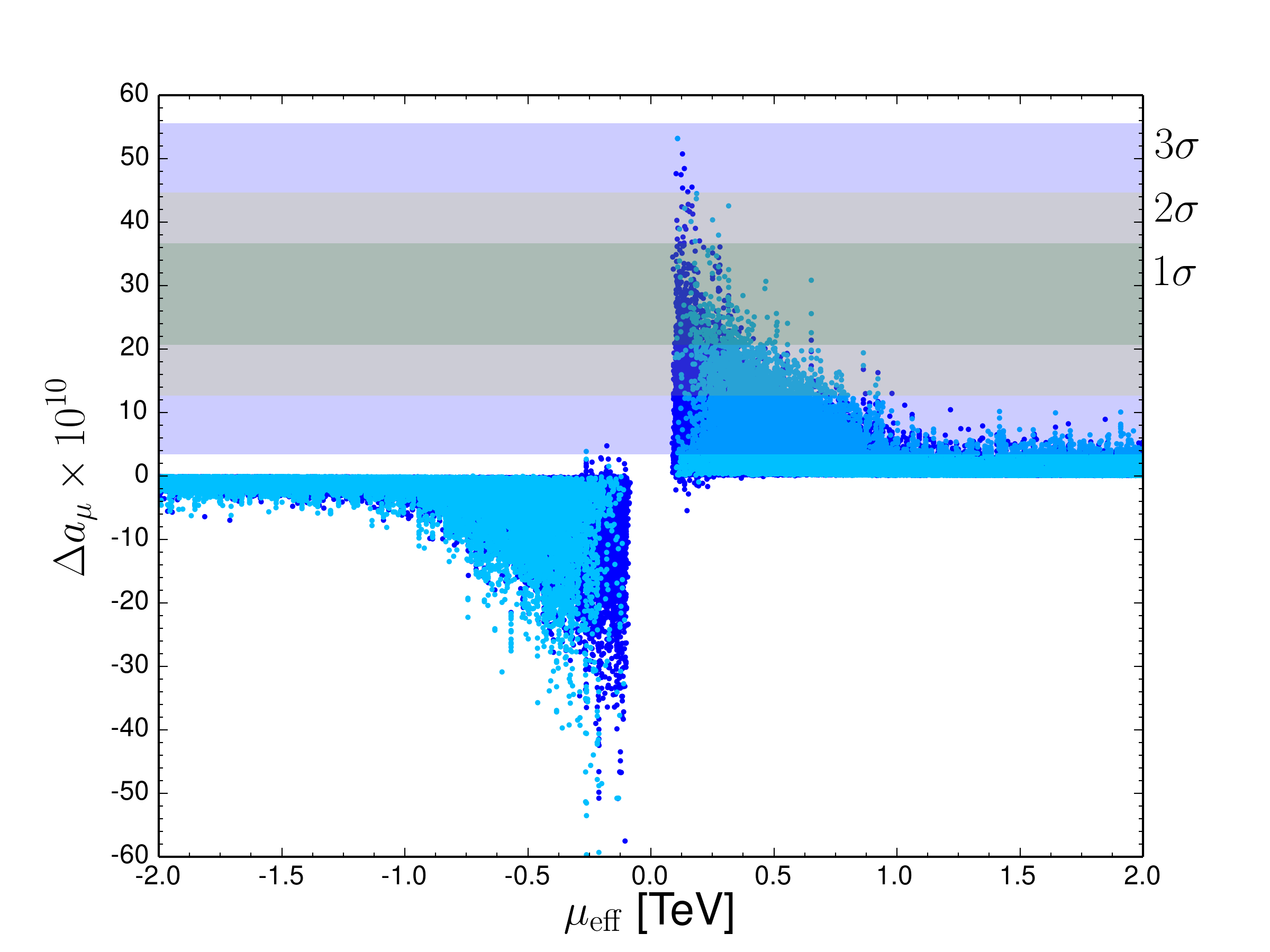}\hspace{.5cm}
  \includegraphics[scale=0.42]{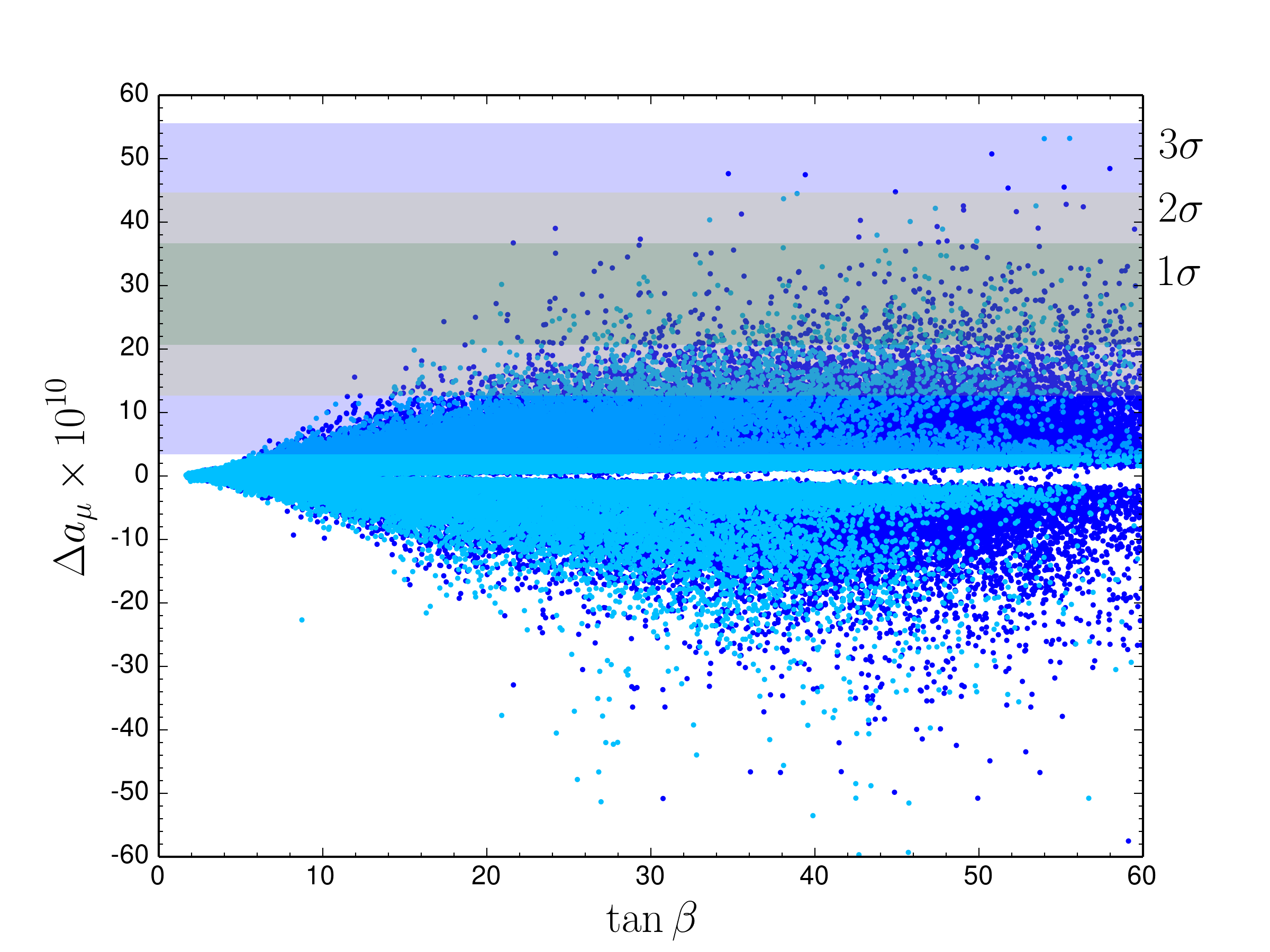}\\
  \includegraphics[scale=0.42]{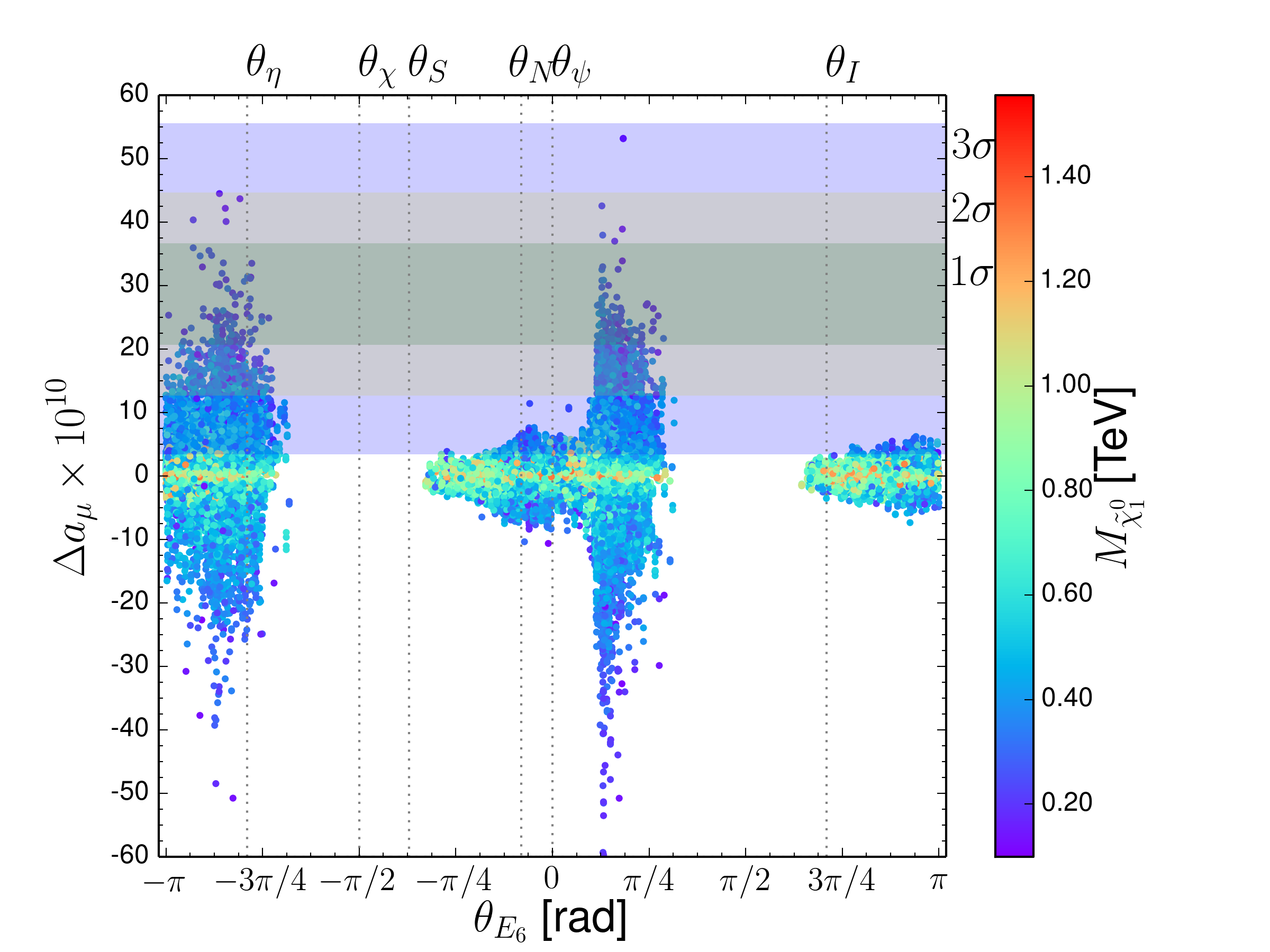}\hspace{0.5cm}
  \includegraphics[scale=0.42]{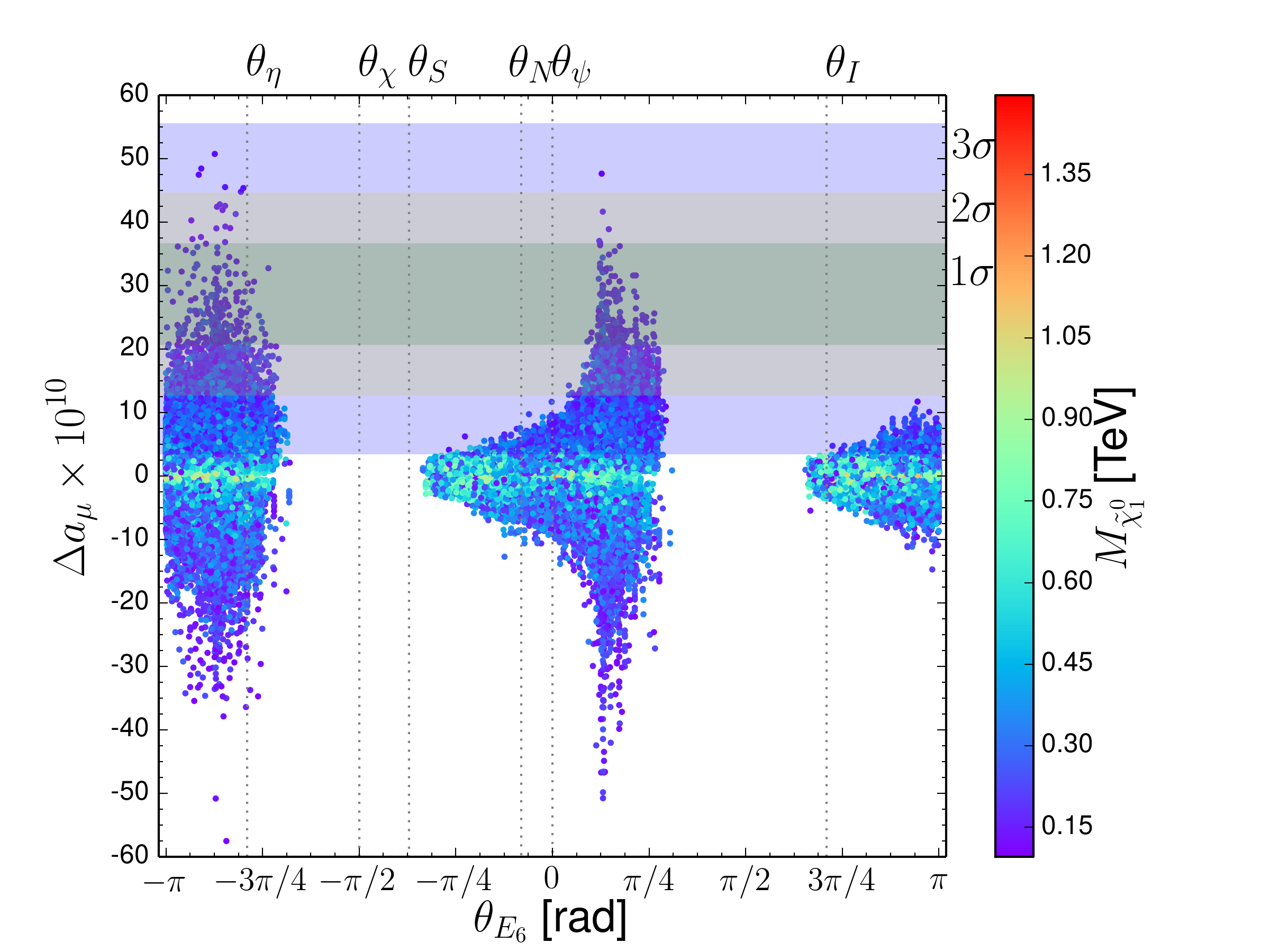}
  \caption{\it UMSSM contributions to the anomalous magnetic moment of the muon,
    $\Delta a_\mu$ shown as a function of the effective $\mu_{\rm eff}$
    parameter (upper left) and $\tan\beta$ (upper right).  The light (dark) blue
    points represent scenarios in which the lightest sneutrino (neutralino) is
    the LSP. On the lower panels of the figure, we present the $\theta_{E_6}$
    dependence of $\Delta a_\mu$ and depict by a colour code the mass of the
    lightest neutralino  for scenarios with a sneutrino (lower left panel) and
    with a neutralino (lower right panel) LSP. On all
    figures, we moreover indicate by a green, grey and purple band the
    $\Delta a_\mu$ values for which we get an agreement at the $1\sigma$,
    $2\sigma$ and $3\sigma$ level with the experimental value, respectively.}
  \label{fig:g-2thE6_tanB}
\end{figure}

For each point of our parameter space scan, we present in the upper panel of
Fig.~\ref{fig:g-2thE6_tanB} the UMSSM contributions to $(g-2)_\mu$, that we
denote by $\Delta a_\mu$, and that is expected to fill the gap between the
theoretical predictions and $(g-2)_\mu$ data. The dependence of $\Delta a_\mu$
on the
$\mu_{\rm eff}$ parameter is depicted on the left panel of the figure, and we
observe that the gap between the experimental measurement and the theoretical
prediction can only be filled for positive value of $\mu_{\rm eff}$. As in the
MSSM, this originates from neutralino and slepton loop contributions that are
proportional to $\mu_{\rm eff}$, so that a negative $\mu_{\rm eff}$ value would
increase and not decrease the discrepancy between theory and experiment.
Sneutrino LSP scenarios mostly feature a small $\mu_{\rm eff}$ value, as already
found in Fig~\ref{fig:MCMC1}, which implies that the lightest neutralino is in
general not too heavy. As a consequence, the corresponding contributions to
$(g-2)_\mu$ are sizable and theoretical predictions agree better with data (for
cases where $\mu_{\rm eff} > 0$). This agreement is in addition facilitated for
large $\tan\beta$ values, as shown in the right panel of the figure. Neutralino
LSP scenarios in contrast allow for intermediate $\mu_{\rm eff}$ values, so
that resulting $\Delta a_\mu$ new physics contributions are not large enough
to entirely fill the experiment-theory gap due to a heavy neutralino mass
suppression.

These conclusions are further confirmed by the lower panel of
Fig.~\ref{fig:g-2thE6_tanB} in which we show the variation of $\Delta a_\mu$ as
a function of the $U(1)'$ mixing angle $\theta_{E_6}$ and correlate the results
with the value of the mass of the lightest neutralino for sneutrino LSP
scenarios (left panel) and neutralino LSP scenarios (right panel). We observe
that in contrast to the other models, $U(1)^\prime_I$ scenarios are unable to
provide an explanation for the $(g-2)_\mu$ observations. This is connected to
the larger $M_{1/2}$ mass parameter typical of these scenarios. The
contributions from $U(1)^\prime$ supersymmetric models to $\Delta a_\mu$ are
dominated by slepton-neutralino loop diagrams, and are maximal for light
sleptons. This occurs when the $D$-terms proportional to $Q_l$ in the slepton
mass matrix are zero as in Fig.~\ref{fig:U1charge}, which corresponds to the
peaks appearing in the lower panel graphs of Fig.~\ref{fig:g-2thE6_tanB}.


\section {Sneutrino Dark Matter}
\label{sec:sneutrinodm}
In this section we concentrate on scenarios exhibiting a sneutrino LSP and
show that sneutrinos are UMSSM viable dark matter candidates, in contrast to the
MSSM possibly extended with right sneutrinos. Unlike in a theory featuring only
the SM gauge group, right sneutrinos can reach, in the UMSSM, thermal
equilibrium thanks to their $U(1)^\prime$ interactions with extra vector and/or
scalar fields. Moreover, the sneutrino pair annihilation cross section is
possibly enhanced by $s$-channel resonant (or
near-resonant) exchanges, and the elastic scattering cross section of a dark
matter particle with a SM parton is suppressed by several orders of magnitude as
sneutrino couplings to the SM $Z$ and Higgs bosons are reduced and the
would-be dominant $Z^\prime$ exchange is mass-suppressed.

Our thorough investigation of the MSSM parameter space has revealed that, when
allowing the model parameters to be small and run freely, the lightest
neutralino naturally emerges as the LSP. Requiring a sneutrino to be the LSP
implies more specific and less general corners of the parameter space, which is
not necessarily an issue as the absence of any beyond the SM signal at the LHC
could be an indication for a non-natural new physics setup. We now focus on the
dark matter implications for all scanned scenarios exhibiting a sneutrino LSP in
the $U(1)^\prime_\psi$, $U(1)^\prime_\eta$ and $U(1)^\prime_I$ models, that
are the three-anomaly free UMSSM setups satisfying so far all current
constraints, and investigate constraints originating from the dark matter relic
abundance in Sec.~\ref{subsec:relic} and direct detection and neutrino fluxes in
Sec.~\ref{subsec:DM_detection}.

\subsection{Relic Density}
\label{subsec:relic}

In order to analyse the constraints that could originate from the relic density
on the UMSSM models, we explicitly choose two possibilities for the $Z'$-boson
mass, a light $Z^\prime$-boson case with $M_{Z^\prime} = 2$~TeV and a heavier
$Z^\prime$-boson case with $M_{Z^\prime} = 2.5$~TeV. Although the former option
is slightly less than the $Z^\prime$-boson limits presented in the 2016 Particle Data Group
review~\cite{Olive:2016xmw}, we recall that such light extra bosons are allowed
in UMSSM scenarios where $Z'$ decays into pairs of supersymmetric particles
contribute significantly. Moreover, we use the
results of
our scan to enforce the values for other parameters to lead to a viable Higgs
boson mass and a fair agreement with all other experiment constraints.
The relic density contributions stemming from
the presence of a $Z'$ boson are crucial for models such as the UMSSM where the
field content of the theory includes right sneutrinos that are not sensitive to
the SM gauge interactions. Whilst a full parameter space scan could be in order,
the above procedure allows us to study and understand the impact of specific
parameters on the relic density, and in particular of the effective
$\mu_{\rm eff}$ parameter and the trilinear coupling $A_\lambda$, as in general,
sneutrino DM is usually overabundant as a result of an inefficient sneutrino
annihilation mechanism. We use as experimental bounds for the relic density the
conservative range provided from the older WMAP data~\cite{Komatsu:2010fb,%
Spergel:2006hy} and including a 20\% uncertainty,
\be
  \Omega_{\rm DM} h^2 = 0.111^{+0.011}_{-0.015} \ .
\ee

\begin{figure}
  \centering
  \includegraphics[scale=0.45]{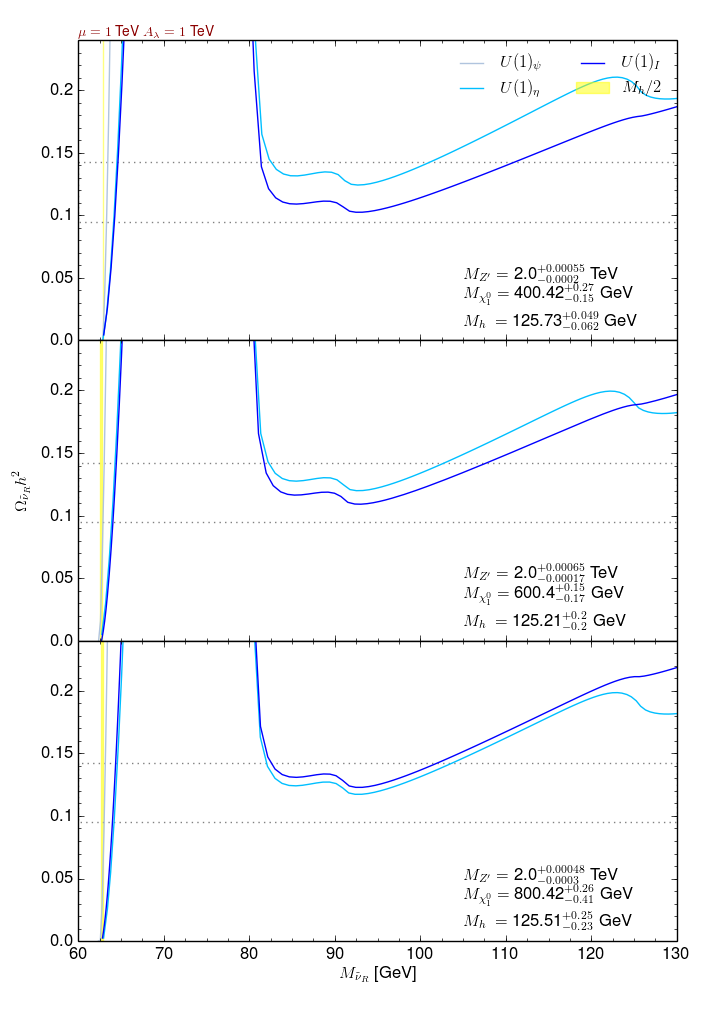}\hspace{0.5cm}
  \includegraphics[scale=0.45]{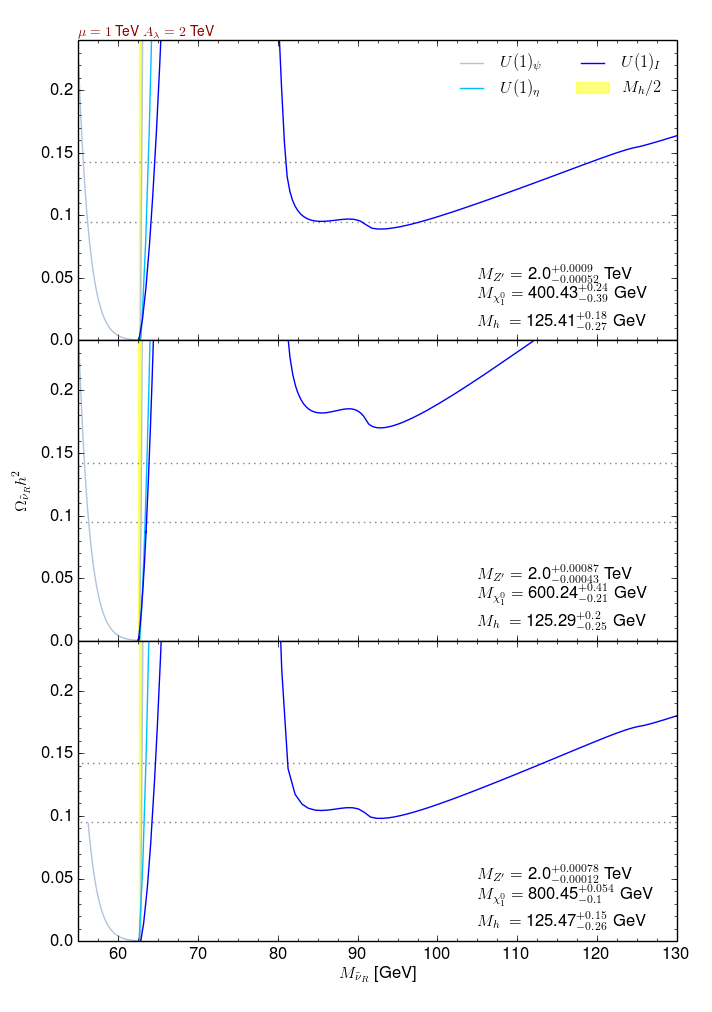}\\
  \includegraphics[scale=0.45]{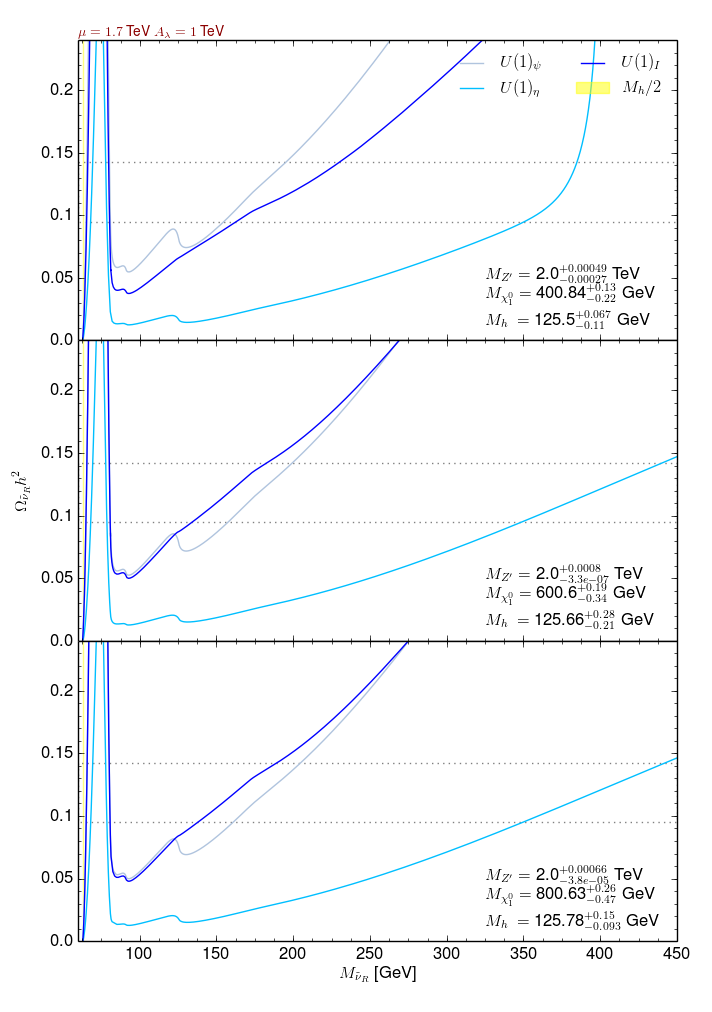}\hspace{0.5cm}
  \includegraphics[scale=0.45]{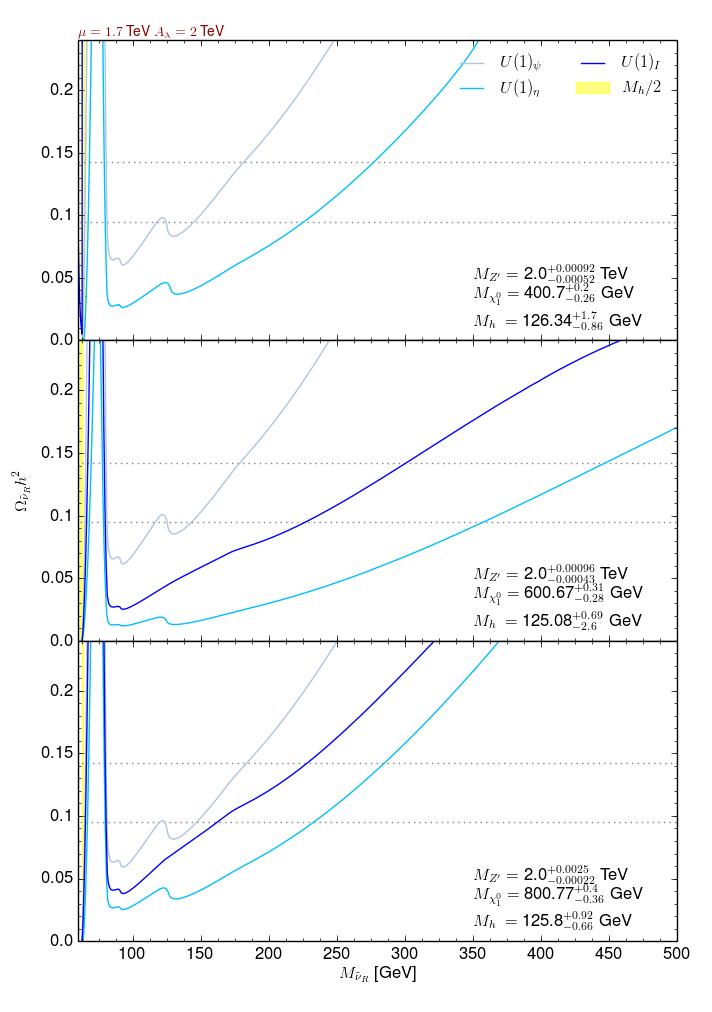}
  \caption{\it Dependence of the relic density for UMSSM scenarios featuring a
    right sneutrino LSP and $M_{Z^\prime}=2$~TeV. We fix $\mu_{\rm eff}$ to
    1~TeV (upper panels) and 1.7~TeV (lower panels), as well as $A_\lambda$ to
    1~TeV (left panels) and 2~TeV (right panels). In each of the four figures,
    the lightest neutralino mass has been respectively fixed to 400~GeV (upper
    inset), 600~GeV (middle inset) and 800~GeV (lower inset) and we focus on the
    the $U(1)^\prime_\psi$ (grey), $U(1)^\prime_\eta$ (light blue) and
    $U(1)^\prime_I$ (dark blue) models.}
  \label{fig:lowZrelic}
\end{figure}

Fixing first $M_{Z^\prime}$ to 2~TeV, we investigate the dependence of the relic
density on the mass of the lightest sneutrino, after selecting varied choices of
$M_{\tilde\chi^0_1}$, $\mu_{\rm eff}$ and $A_\lambda$. In addition, the
$\tan\beta$, $M_0$ and $A_0$ parameters are modified correspondingly to recover
a correct lightest Higgs boson mass of 125~GeV, and agreement with all the
previously discussed experimental constraints. We consider, in our analysis,
three parameter space regions on the basis of the mass of the lightest
neutralino, which is most often the
next-to-lightest superpartner (NLSP), here taken to be 400~GeV, 600~GeV
and 800~GeV respectively. Alongside the neutralino mass, $\mu_{\rm eff}$ and
$A_\lambda$ are set to 1 or 1.7~TeV and 1 or 2~TeV respectively, this restricted
set of values being sufficient to investigate the effects on these parameters on
the dark matter relic density. The results are presented in
Fig.~\ref{fig:lowZrelic}.

\begin{figure}
  \centering
  \includegraphics[scale=0.45]{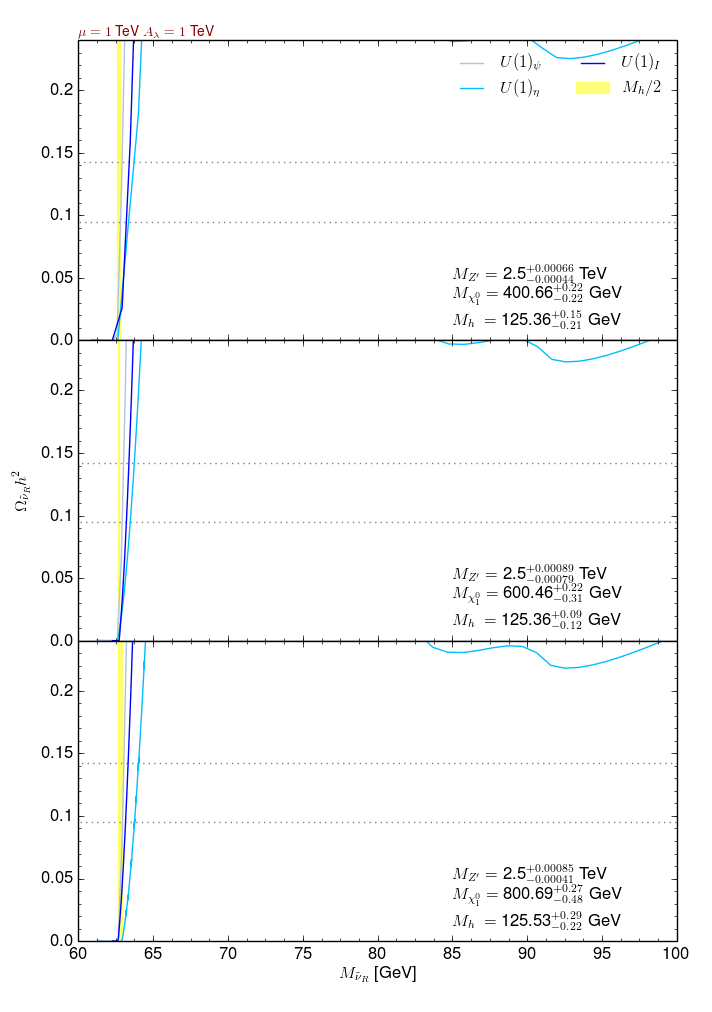}\hspace{0.5cm}
  \includegraphics[scale=0.45]{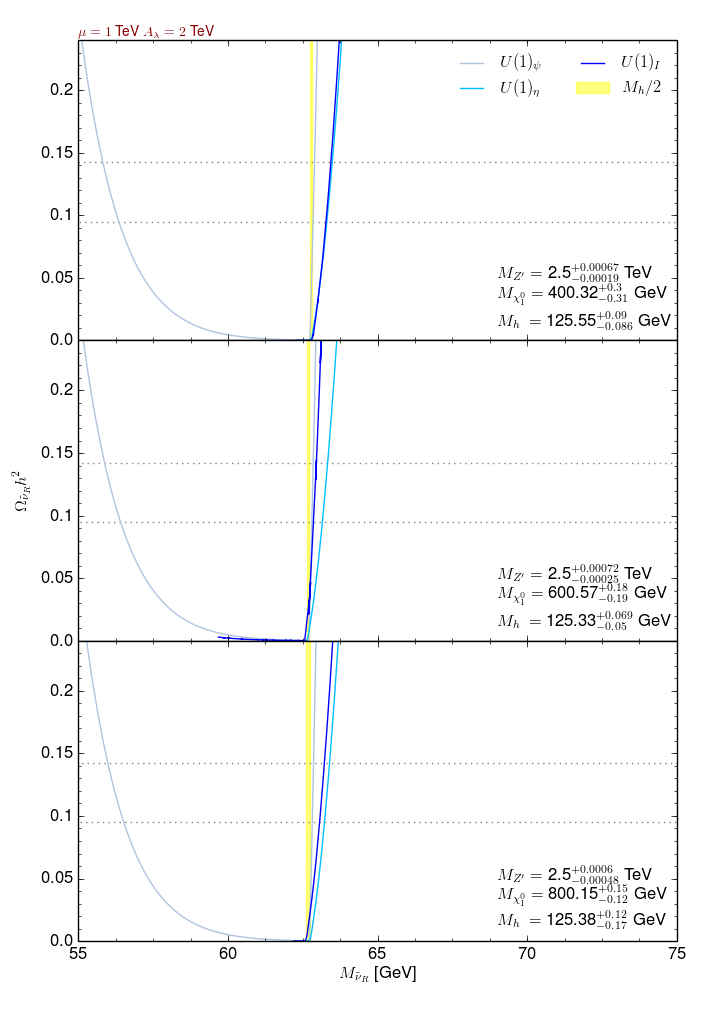}\\
  \includegraphics[scale=0.45]{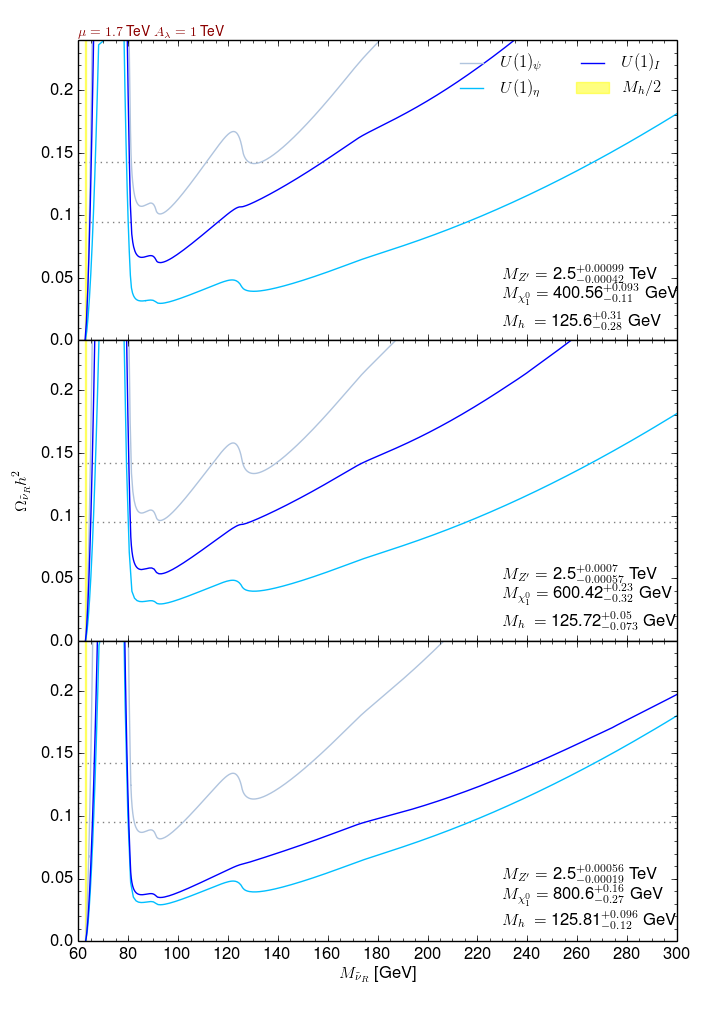}\hspace{0.5cm}
  \includegraphics[scale=0.45]{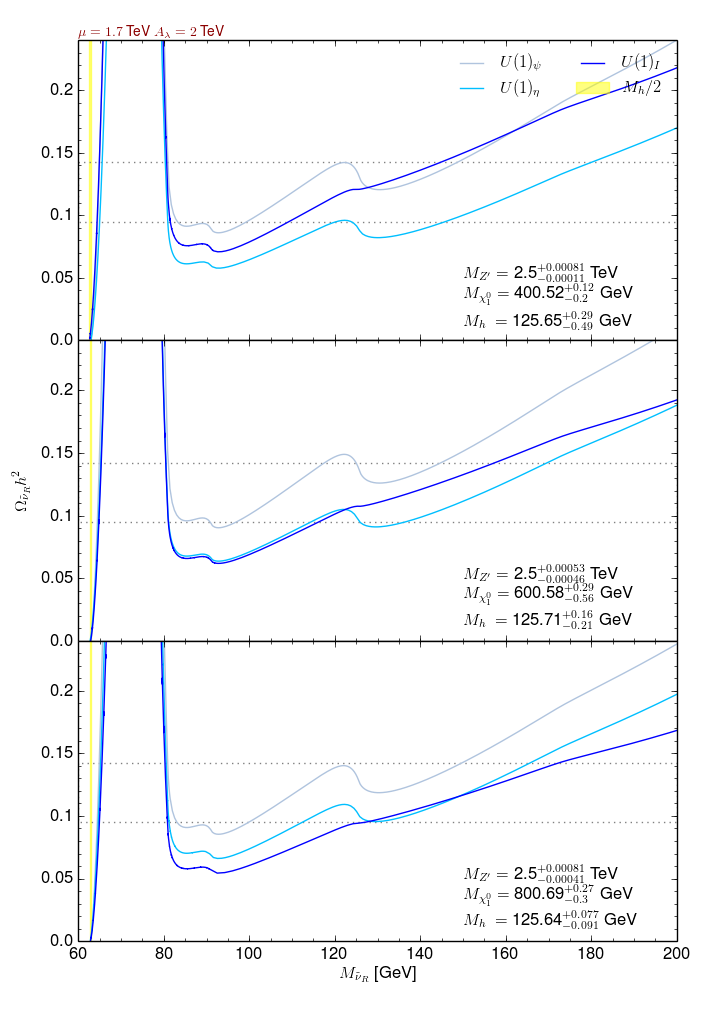}
  \caption{\it Same as in Fig.~\ref{fig:lowZrelic} but for $M_{Z'}=2.5$~TeV.}
  \label{fig:midZrelic}
\end{figure}

In the upper right panel of the figure, we set $\mu_{\rm eff}=A_\lambda=1$~TeV
and show that regardless of the value of the lightest neutralino mass and
depending on the class of $U(1)'$ model, there exist two regimes where the
predicted relic density matches the observations. First, in a region where the
sneutrino mass is close to 65~GeV, one can design $U(1)^\prime_\eta$,
$U(1)^\prime_\psi$ and $U(1)^\prime_I$ UMSSM models where the relic density
bounds are
satisfied. A correct dark matter annihilation cross section can be achieved
thanks to the enhanced contributions of Higgs-boson exchange diagrams that
proceed in a
resonant or near-resonant production mode ($M_{\tilde\nu_1} \approx M_h/2$).
This configuration, also known as a Higgs funnel configuration, is achievable
for any value of the $\mu_{\rm eff}$ and $A_\lambda$ parameters, as shown in the
other panels of Fig.~\ref{fig:lowZrelic}, although the value of $A_\lambda$
affects its size. The Higgs funnel region is indeed narrower for larger
$A_\lambda$ values. While a similar regime could be expected for
$M_{\tilde\nu_1} \approx M_{Z'}/2$, this latter setup implies very heavy
sneutrinos that are then incompatible with the requirement of a sneutrino
being the LSP.

A second kinematical regime allows for the recovery of a proper relic density,
with a sneutrino mass lying in the $[80, 110]$~GeV window for $\mu_{\rm eff}=
A_\lambda=1$~TeV (upper left panel of the figure). In this regime, both dark
matter annihilation into a pair of $Z$-bosons and LSP-NLSP co-annihilations are
important, as noticed by the size of the region depending on the mass of the
lightest neutralino. Investigating the other panels of the figure, one observes
that the exact details of this region of the parameter space, as well as its
existence, strongly depend on the values of the $\mu_{\rm eff}$ and $A_\lambda$
parameters. The latter indeed directly affect the nature of the lightest
neutralino and the properties of the heavier part of the Higgs sector, $h_1$
exchange contributions being very relevant for a not too heavy next-to-lightest
Higgs boson (see Fig.~\ref{fig:MCMC3}).

$Z'$-boson exchange contributions play nevertheless a key role in the
calculation of the relic density. For instance, in $U(1)^\prime_\psi$ scenarios,
the new gauge interactions of the sneutrinos are relatively weaker (due to the
involved $U(1)'$ charges), the corresponding branching ratio being three times
smaller than for the two other cases. As a result, the existence of the heavier
sneutrino regime itself, in which the relic density constraints are correctly
satisfied, is more challenging. This feature is emphasised on
Fig.~\ref{fig:midZrelic} where the $Z'$-boson mass is pushed to 2.5~TeV, the
other $M_{\tilde\chi_1^0}$, $A_\lambda$ and $\mu_{\rm eff}$ parameters being
varied as before whereas the $\tan\beta$, $M_0$ and $A_0$ parameters are once
again adjusted to reproduce all previously considered constraints. Although the
existence of the Higgs funnel regime is barely affected by the changes, this
regime may be shifted towards lighter sneutrino masses in the $[50, 65]$~GeV
regime. In addition, heavier sneutrino LSP scenarios are more difficult to
accommodate, which directly prevents the heavy sneutrino regime with a
consistent relic density from existing, in particular if the $\mu_{\rm eff}$
parameter is not large enough.

\subsection{Constraints from dark matter direct detection and neutrino fluxes}
\label{subsec:DM_detection}

\begin{figure}
  \centering
  \includegraphics[scale=0.45]{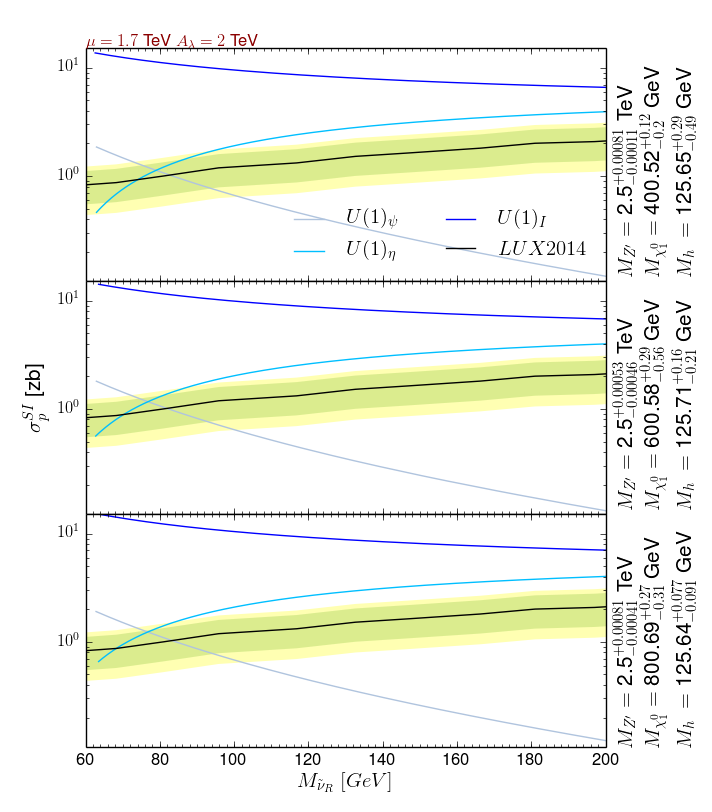}
  \includegraphics[scale=0.45]{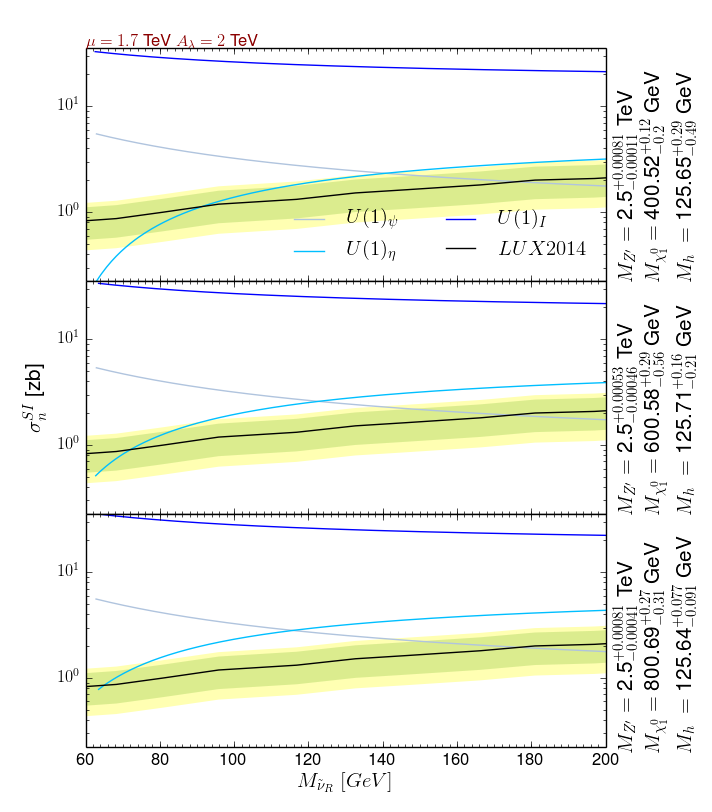}
  \caption{\it Spin independent cross section associated with the scattering of
   dark matter off protons (left) and neutrons (right) presented as functions of
   the dark matter mass. We fix $\mu_{\rm eff}$ to 1.7~TeV and $A_\lambda$ to
   2~TeV. In each of the subfigures, the lightest neutralino mass has been
   respectively fixed to 400~GeV (upper inset), 600~GeV (middle inset) and
   800~GeV (lower inset) and we focus on the $U(1)^\prime_\psi$ (grey),
   $U(1)^\prime_\eta$ (light blue) and $U(1)^\prime_I$ (dark blue) models. The
   band corresponds to the $2\sigma$ limits extracted from LUX
   data~\cite{Akerib:2015rjg,mark_mitchell_2016_61108}.\label{fig:SI}}
  \includegraphics[scale=0.52]{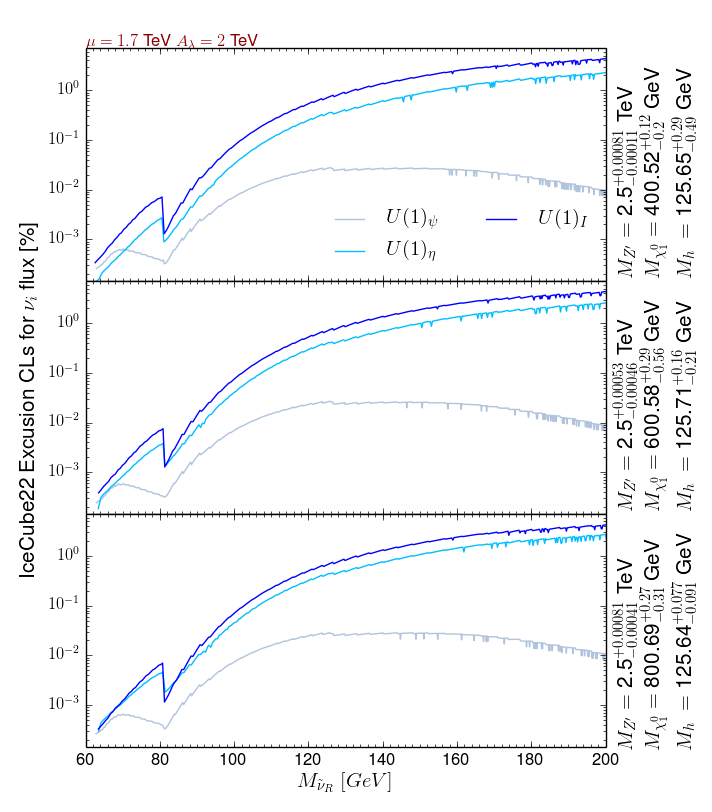}
  \caption{\it Exclusion bounds, given as a confidence level, extracted from
    the neutrino flux observed in the IceCube experiment and presented as a
    function of the lightest sneutrino mass. The UMSSM scenario is fixed as in
    Fig.~\ref{fig:SI}. \label{fig:IceCube}}
\end{figure}

Direct detection experiments aim to detect DM scattering off nuclear matter and
to measure its properties. While the DM interactions with
nuclear matter can be generally classified as spin-dependent or
spin-independent, only the latter is relevant for sneutrino dark matter. We
present, in Fig.~\ref{fig:SI}, UMSSM predictions for the spin-independent cross
section associated with the scattering the LSP with protons (left panel) and
neutrons (right panel), and compare them with the experimental results from the
LUX experiment~\cite{Akerib:2015rjg}. We
adopt UMSSM scenarios in which $\mu_{\rm eff}= 1.7$~TeV and $A_\lambda=2$~TeV,
and the $Z'$ mass is fixed to 2.5~TeV. As in the previous section, the results
are given for lightest neutralino masses of 400~GeV (top inset), 600~GeV
(central inset) and 800~GeV (lower inset).

Our results demonstrate the discriminating power of the spin-independent
DM-nucleon scattering cross section as its behaviour as a function of the mass
of the sneutrino LSP highly depends on the $U(1)'$ model. For a given LSP mass,
cross section values obtained in $U(1)'_I$ models are one order of magnitude
larger than for the two other classes of models, $U(1)'_\eta$ cross sections
increasing in addition with the sneutrino mass. The results of the LUX
experiment introduce strong constraints on wide regions of the parameter space,
and our specific $\mu_{\rm eff}$
and $A_\lambda$ choice are typical from the parameter space region in which both
the relic density and the direct detection constraints can be easily
accommodated. This however introduces tensions with the parameter space regions
favoured by the anomalous magnetic moment of the muon results (see
Sec.~\ref{subsec:muon_g-2}), and only the Higgs funnel region in which the
sneutrino mass is half of the Higgs-boson mass survives too all constraints.

While $U(1)'_I$ models are clearly disfavoured by direct detection data,
$U(1)^\prime_\psi$ scenarios cannot feature a viable light sneutrino DM, whilst
$U(1)'_\eta$ setups in contrast prefer light LSP configurations with a sneutrino
mass of about 60~GeV to 100~GeV depending on the neutralino mass.
These results stem from the interaction of the lightest sneutrino with the $Z$,
$Z'$ and Higgs bosons. As the lightest sneutrino only very weakly couples to the
SM sector, the scattering cross section mostly depends on the vectorial
couplings of the LSP and of the SM quarks to the $Z'$-boson. These quark
vectorial couplings being vanishing in the $U(1)'_\psi$ model, the resulting
cross section is largely suppressed and those scenarios can survive more
easily to LUX data. The neutron cross section, being larger as
expected~\cite{Belanger:2015cra}, however drastically reduces the size of the
allowed region of the parameter space and future improvements in direct
detection experiments may directly challenge the studied UMSSM setups.

Recent observations of ultra-high energy neutrino events at the IceCube
experiment~\cite{Abbasi:2009uz} indicate a possible deficit in the amount of
observed muon tracks, which is known as the muon deficit problem, and an
apparent energy gap in the three-year high energy neutrino data. This challenges
any explanation based on atmospheric neutrinos, and suggests an
extra-terrestrial origin that could involve dark matter. Data being however
consistent with the SM expectation, this may introduce extra constraints when DM
model building is at stake. We present, in Fig.~\ref{fig:IceCube}, the
corresponding exclusion as obtained in the UMSSM setup considered in this
section with the help of \textsc{micrOMEGAs}. This shows
that even if genuine differences amongst the three considered $U(1)'$ options
once again appear, in particular for large sneutrino masses, all results are
consistent with the SM to a good extent.


\section{Neutralino dark matter}
\label{sec:neutralinoLSP}

\begin{figure}
  \centering
  \includegraphics[scale=0.42]{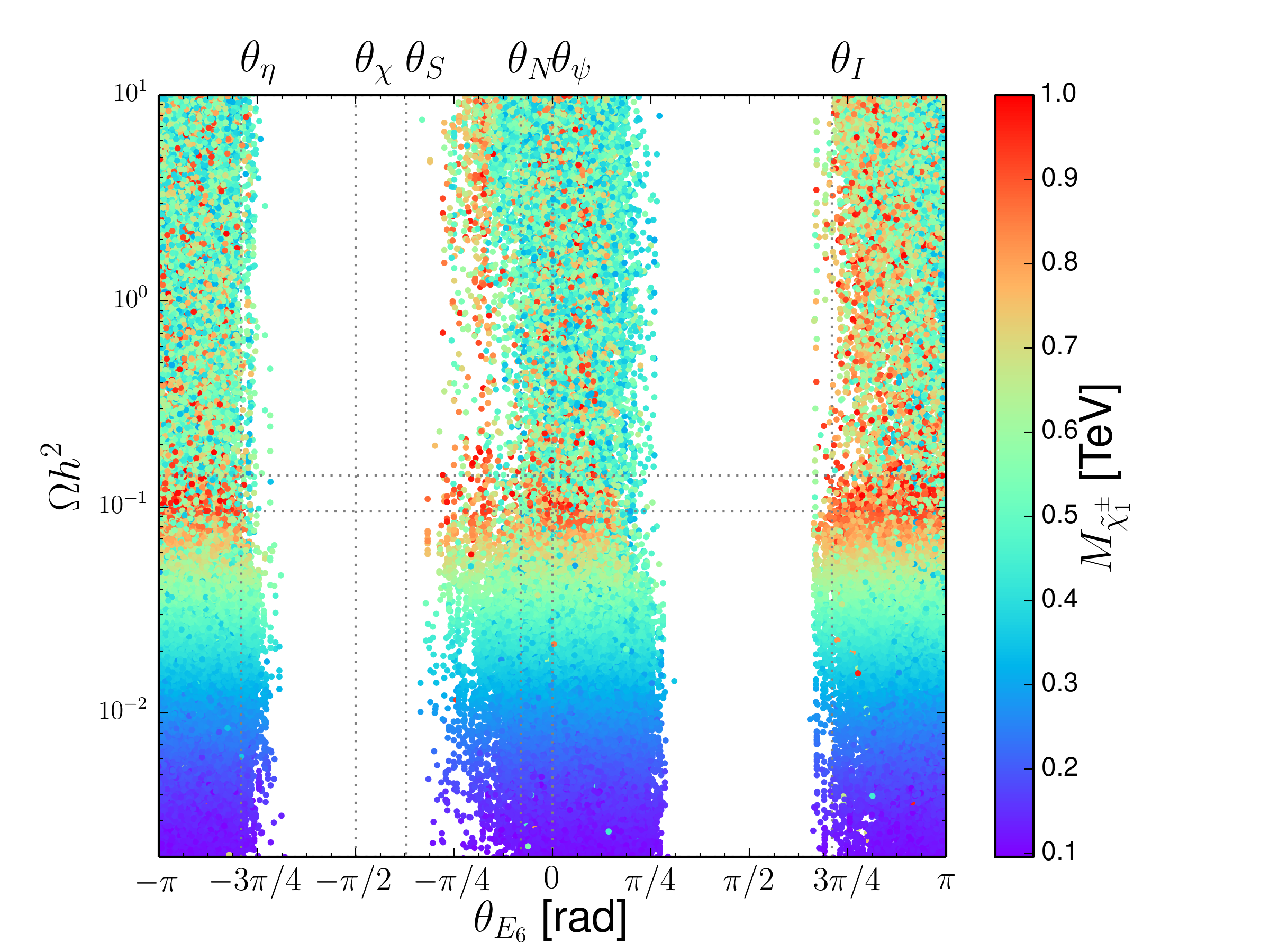}\hspace{0.5cm}
  \includegraphics[scale=0.42]{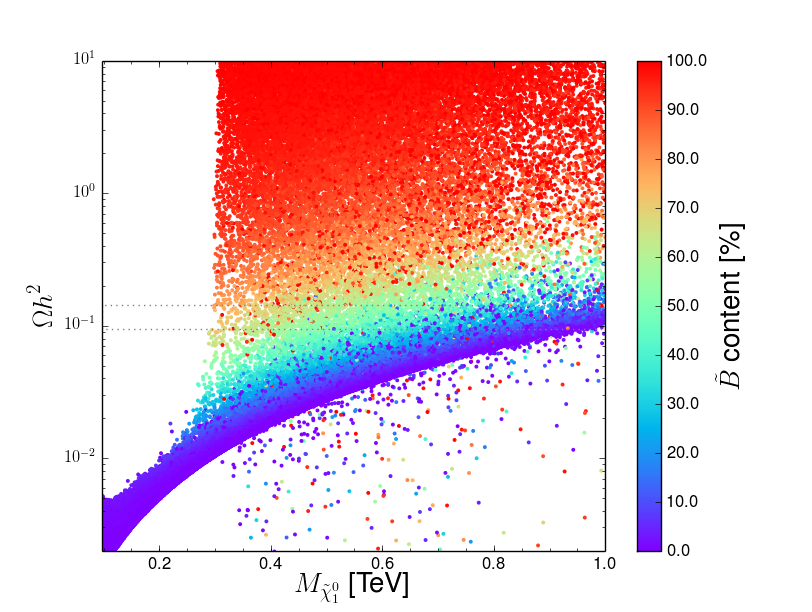}\\
  \includegraphics[scale=0.42]{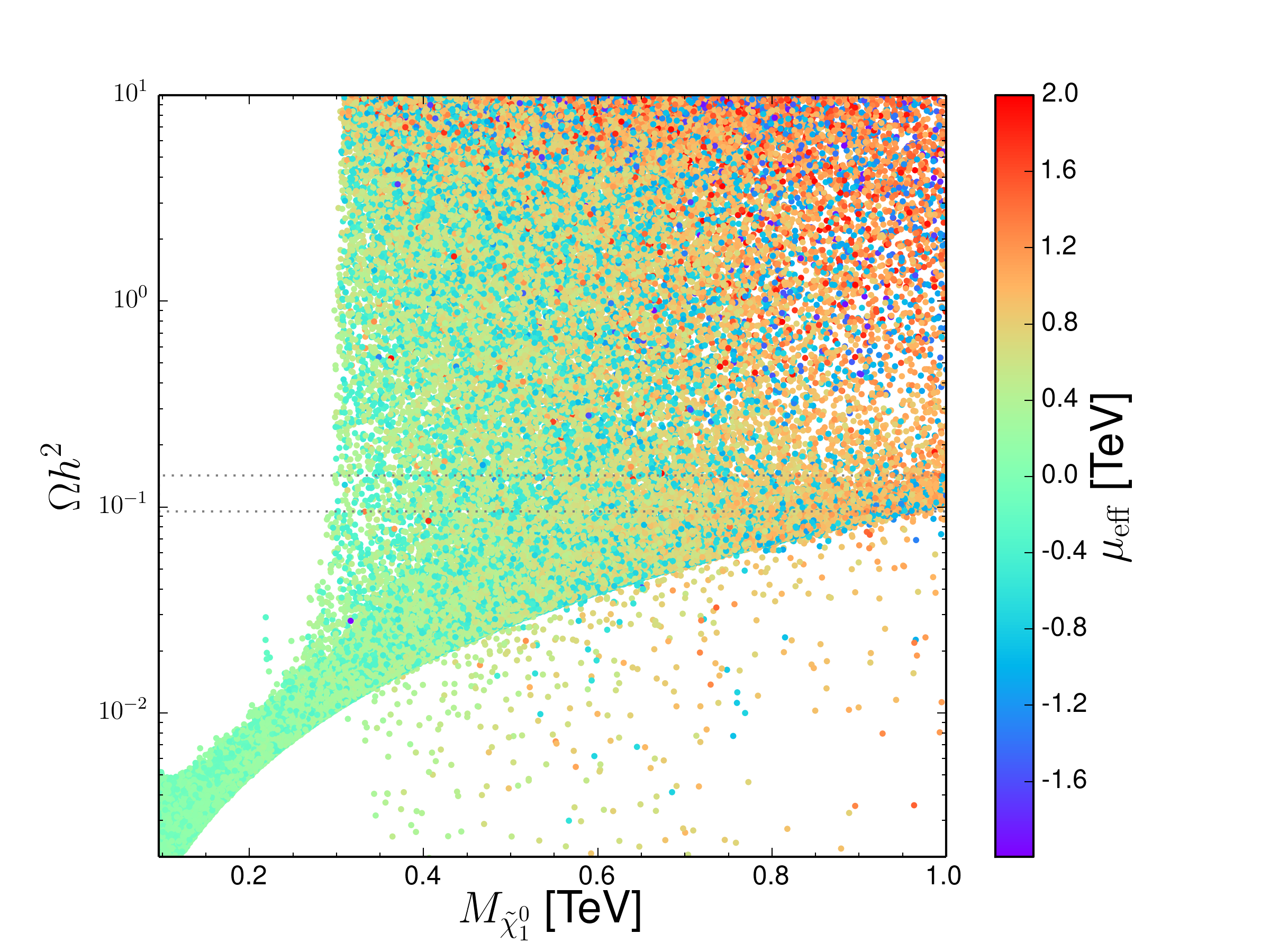}\hspace{0.5cm}
  \includegraphics[scale=0.42]{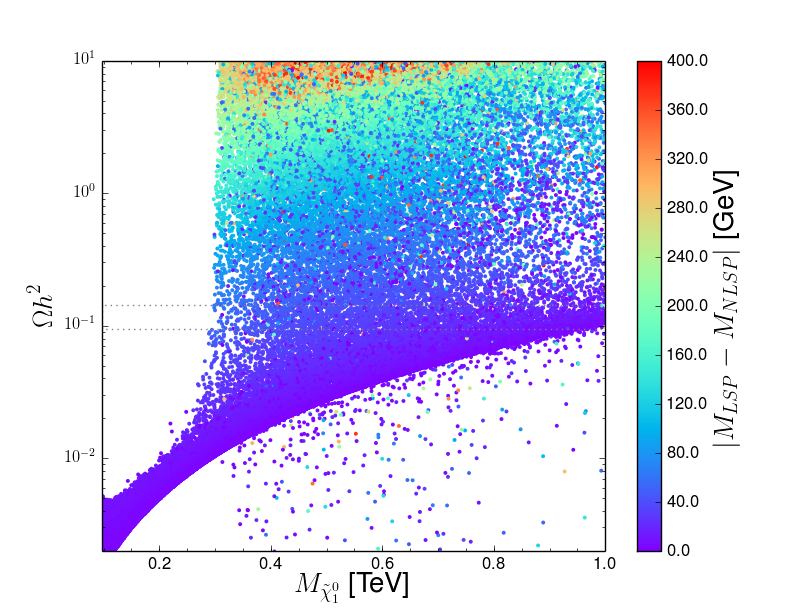}
  \caption{\it DM relic density obtained for UMSSM scenarios featuring a
   neutralino LSP, presented as a function of the LSP mass and the $U(1)'$
   mixing angle (upper left panel), the neutralino bino component (upper right
   panel), the $\mu_{\rm eff}$ parameter (lower left panel) and the mass
   difference between the LSP and the NLSP (lower right panel).}
  \label{fig:neutralinorelic}
\end{figure}

As shown in the above sections, the LSP can naturally be the lightest
neutralino, that consists in UMSSM scenarios of
an admixture of $\lambda_{\tilde B}$, $\lambda_{\tilde W}$ and
$\lambda_{\tilde {Z}^\prime}$ gauginos, as well as of higgsinos. Whether the
LSP in a particular setup is able to yield the right relic abundance
depends crucially on its composition. For a bino-dominated or a
bino$'$-dominated neutralino, the LSP is a gauge singlet and it annihilates
mainly through sfermion $t$-channel exchanges. As sfermions are heavy, the
annihilation mechanism is often inefficient so that accommodating the observed
relic density is difficult, unless one strongly relies on co-annihilations. The
relic density can be more easily reproduced when the lightest neutralino is of a
higgsino or wino nature, or a mixed state. If the LSP is higgsino-like, its
mass is driven by the $\mu_{\rm eff}$ parameter, as is the mass of the
lightest chargino and of the next-to-lightest neutralino. These three particles
being almost degenerate, annihilations and co-annihilations easily occur so
that DM could be underabundant if the LSP is too light~\cite{Baer:2011ab}. In
our setup the wino-like LSP is in contrast impossible to be realised due to the
GUT relations that we have imposed in our scanning procedure.

Unlike for sneutrinos, the neutralino LSP mass is mainly determined by the
$M_{1/2}$ and the $\mu_{\rm eff}$ parameters that also affect {\it all} the
particle masses of the model. The LSP mass cannot be consequently varied
independently of the rest of the spectrum, making an analysis based on specific
benchmark configurations less straightforward than in the sneutrino LSP case. We
therefore base our study on the results of our parameter space scan where all
the constraints described in Sec.~\ref{sec:constr} are imposed. Our results are
given in Fig.~\ref{fig:neutralinorelic} where we present the dependence of the
DM relic density on the LSP mass. We correlate our findings with the value of
the $U(1)'$ mixing angle $\theta_{E_6}$ (upper left panel), the magnitude of the
LSP bino component (upper right panel), the value of the $\mu_{\rm eff}$
parameter (lower left panel) and the mass difference between the LSP and the
NLSP (lower right panel). Accommodating the correct relic density yields a LSP
mass of at least 300~GeV, which contrasts with sneutrino LSP scenarios where the
mass of the latter is smaller. As expected, the lightest neutralino is mostly
bino-like, and a higgsino component is only allowed for heavier LSP setups so
that the co-annihilation rate turns out to be tamed. Viable DM scenarios also
feature a small $\mu_{\rm eff}$ parameter lying in the $[-400, 400]$~GeV mass
window, which allows the next-to-lightest neutralino to be higgsino-like and not
too heavy, as emphasised in the lower right panel of the figure as it is often
the NLSP. Co-annihilations
are hence under good control, which guarantees a relic density in agreement
with the observations. Our results also show that small differences are
present for the different $U(1)'$ scenarios under consideration, the LSP mass
being only in general slightly larger for $U(1)^\prime_I$ models.

\begin{figure}
  \centering
  \includegraphics[scale=0.42]{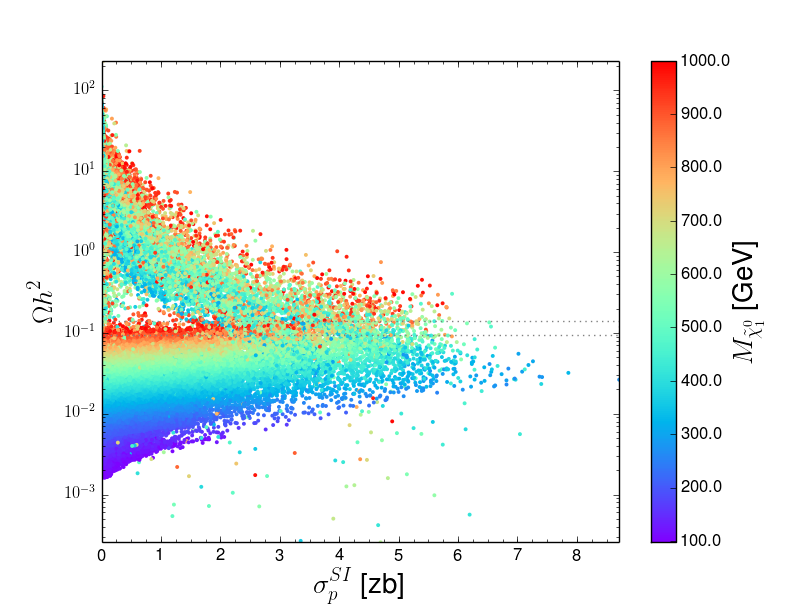}\hspace{0.5cm}
  \includegraphics[scale=0.42]{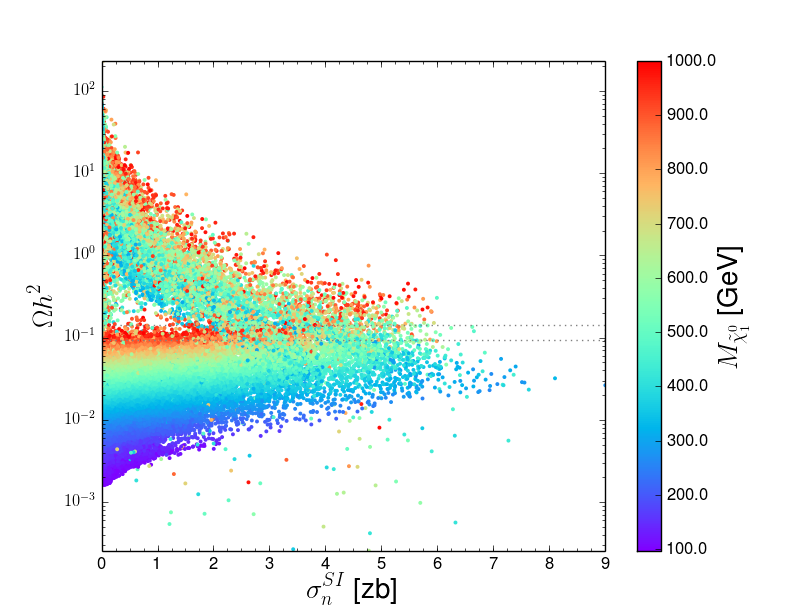}\\
  \includegraphics[scale=0.42]{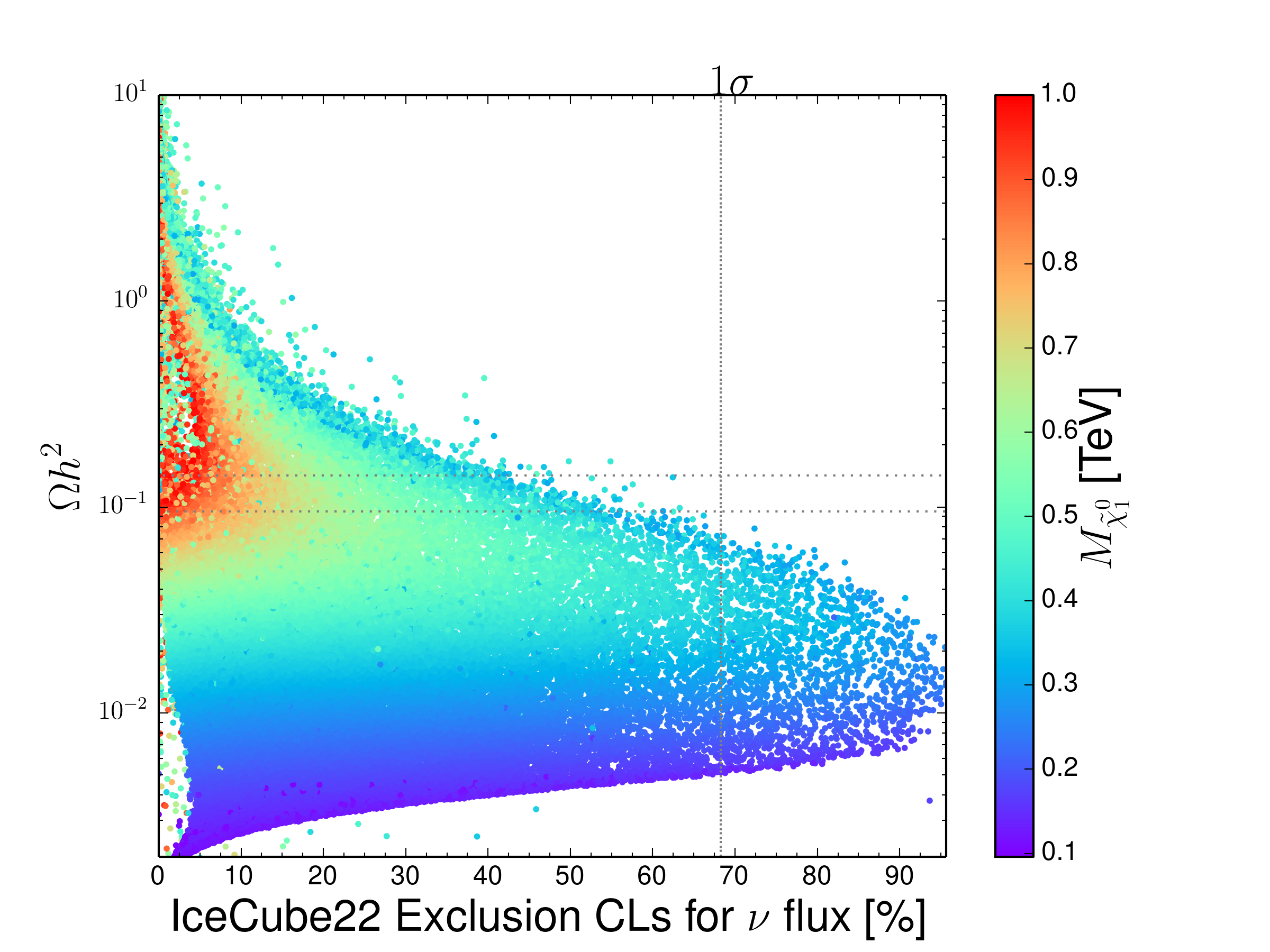}\hspace{0.5cm}
  \includegraphics[scale=0.42]{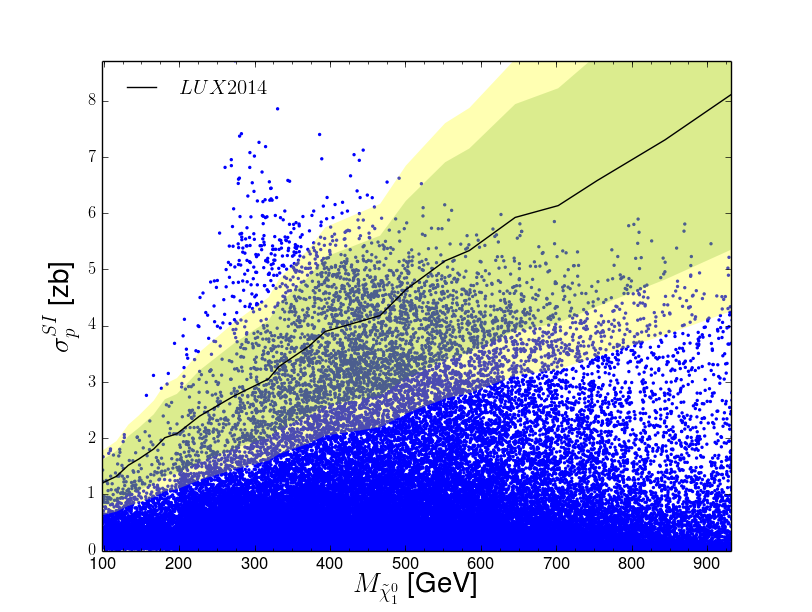}
  \caption{\it Constraints on the UMSSM parameter space region in which the LSP
    is a neutralino that originate from DM direct detection. We present the
    dependence of the relic density on the neutralino mass and on the resulting
    spin-independent dark matter scattering cross section with protons (upper
    left panel) and neutrons (upper right panel) and on the possible exclusion
    that could be obtained from IceCube  results (lower left panel). We also
    show the dependence of the spin-independent DM-proton scattering cross
    section on the neutralino mass, including the bound stemming from the LUX
    experiment (lower right panel).}
  \label{fig:direct_neutralino}
\end{figure}

In Fig.~\ref{fig:direct_neutralino}, we include constraints that arise from DM
direct detection experiments and correlate the proton-DM (upper left panel) and
neutron-DM (upper right panel) spin-independent scattering cross section with
the predicted relic density, including in addition information on the LSP mass
for each point. This shows, together with the results of the lower right panel
of the figure, that regardless
of the LSP mass, there are always scenarios for which both the relic density and
the direct detection constraints can be satisfied. We finally correlate, in the
lower left panel of the figure, the relic density and
the confidence level exclusion that can be obtained from the IceCube results on
the neutrino flux. We observe that contrary to the sneutrino LSP case, here
neutrino flux results play a role in constraining the UMSSM parameter space.

\section{Collider signals}
\label{subsec:benchmarks}

New physics models featuring a dark matter candidate can in general be equally
tested with cosmology and collider probes and extra pieces of information can be
obtained when both sources of constraints are considered in 
complementarily~\cite{Arina:2016cqj}. In the previous sections, we have
discussed the DM phenomenology of UMSSM realisations in which the LSP is either
the lightest sneutrino or the lightest neutralino, with the hope of getting
handles allowing for the distinction of the gauge group structure. In this
section, we focus on the potential searches that could be performed at the
LHC, in particular when a part of the particle spectrum is light and when the
high-luminosity LHC run is considered. To determine the signals to be
searched for, we focus on a set of promising benchmarks obtained from our scan
results for which all constraints are satisfied. This in particular
concerns scenarios featuring a light sneutrino LSP. In order to evaluate the
fiducial cross sections associated with various signals, we export the UMSSM to
the UFO format~\cite{Degrande:2011ua} and make use of the {\textsc MG5\_aMC@NLO}
framework version~2.4.3~\cite{Alwall:2014hca} to simulate hard-scattering LHC
collisions. The QCD environment characteristic of hadronic collisions is
simulated by means of the {\textsc Pythia}~8 program
version~8.2.19~\cite{Sjostrand:2014zea} and we rely on the {\textsc Delphes}~3
package version~3.3.2~\cite{deFavereau:2013fsa} for the modelling of the
response of a typical LHC
detector. The resulting detector-level events are reconstructed by using the
anti-$k_T$ jet algorithm~\cite{Cacciari:2008gp} as embedded in the {\textsc FastJet}
library version 3.1.3~\cite{Cacciari:2011ma}, and the reconstructed events are
analysed within the {\textsc MadAnalysis}~5 framework
version~1.4.18~\cite{Conte:2012fm}.

The best studied DM signatures at the LHC consist of the mono-$X$ probes
for which a certain amount of missing transverse energy (carried by one or more
DM particles) is produced in association with a single energetic visible
SM object. As in the case of other models, monojet signals are thought as the
most promising due to the relative magnitude of the strong coupling with respect
to the other gauge couplings. The corresponding rates are however very reduced
in the case of a sneutrino LSP, in particular once one imposes a typical monojet
selection that requires the presence of a jet with a large transverse momentum
and a veto on final state leptons. Additionally, dark matter can also be
produced
together with an electroweak vector boson or a $Z^\prime$ boson radiated off the
initial state. While the corresponding production cross section is expected to
be smaller than the monojet one, the final state offers more freedom to reject
the background and is thus worthy to be searched for. Moreover, if as in the
UMSSM case, DM particles strongly couple to SM or extra gauge bosons,
mono-vector boson production may be the dominant channel yielding DM production
at the LHC. However, once all the constraints considered in the previous section
are imposed, the remaining regions of the parameter space correspond to cross
sections that are either negligible or too small relatively to the background
cross sections.

Another way to probe phenomenologically viable UMSSM scenarios is to focus on
sfermion pair production, and in particular on the production of the lighter
third generation sfermions. The considered UMSSM scenarios feature heavy stops
and sbottoms, so that third generation squark pair-production could be in
principle easily tagged thanks to the subsequent presence of very hard final
state objects. However, the associated production total rates are of the order
of at most 1~fb. This makes any new physics contribution impossible to observe
relative to the overwhelming SM background, even if advanced analysis
techniques relying on the shape of the differential distributions are used.
Moving on with the slepton sector, stau pair production is not expected to offer
any extra handle on UMSSM-induced new physics, as the related rates are
suppressed due to the electroweak nature of the process. The possible
enhancement arising from the $Z'$ contributions is in addition reduced given the
low $Z'$-bosons branching ratios into sleptons (see Fig.~\ref{fig:ZpBR}).

Finally, we have studied chargino and neutralino pair-production, and in
particular the associated production of one chargino and one neutralino that
could be enhanced when the effective $\mu_{\rm eff}$ parameter is
small~\cite{Frank:2012ne}. The subsequent associated signatures can contain one,
two or more than two leptons, jets and missing energy. Fiducial cross sections
of the order of the fb are obtained, which are nonetheless too small to be
distinguished from the SM background even after relying on a judicious selection
strategy.

The challenges of observing viable UMSSM models at colliders are not unique, and
it turns out that scenarios that are in principle observable at colliders are
disfavoured by cosmology, and that scenarios in agreement with cosmological and
astrophysical data are out of reach of any present collider.


\section{Summary and Conclusion}
\label{sec:conclusion}

We have presented an extensive phenomenological exploration of $U(1)^\prime$
supersymmetric models that can be classified according to the way a grand
unified $E_6$ symmetry would be broken. Our study has revealed that a large
volume of the parameter space is compatible with constraints originating from
cosmology, astrophysics, precision tests, Higgs physics at the LHC,  but that
these constraints have simultaneously a significant impact on the determination
of the favoured regions of the parameter space. As allowed scenarios can equally
feature the lightest sneutrino or the lightest neutralino  as dark matter
candidates, we have investigated the existence of handles to differentiate
between these two options in the context of the five anomaly-free $U(1)^\prime$
setups.  We have scanned the UMSSM parameter space for phenomenologically viable
models, imposing unification conditions at the GUT scale and allowing the
remaining parameters to run freely. While sneutrino LSP is possible only in
$U(1)^\prime_\psi$, $U(1)^\prime_\eta$ and $U(1)^\prime_I$ models,
$U(1)^\prime_N$ setups have the particularity that right sneutrinos decouple
from the $U(1)^\prime$ sector so that only a neutralino LSP can be a viable dark
matter candidate. In addition, anomaly-free $U(1)^\prime_\chi$ and
$U(1)^\prime_S$ realisations cannot induce a viable symmetry breaking pattern
due to the $U(1)^\prime$ charge assignments and are thus excluded.

We have additionally found that in general the neutralino emerges as the most
natural LSP, but that sneutrino LSP scenarios are possible when the
corresponding soft masses are small, which is possible as they are independent
of the other parameters. In particular, the parameter space region consistently
preferred by the cosmology is when $M_{\tilde\nu_1} \simeq M_h/2$. This inhabits
the so-called Higgs-funnel regime where the observed relic density can be
accommodated thanks
to the contributions of resonant Higgs exchange diagrams. All direct detection
bounds can additionally be satisfied, and an explanation for anomalous magnetic
moment of the muon data can be proposed for $U(1)^\prime_\eta$ and
$U(1)^\prime_\psi$ models, all other $U(1)'$ models being unable to fulfill all
constraints at the same time. Depending on other parameter values, higher
sneutrino mass regions can open up, but such regions are limited by other dark
matter constraints. For the considered benchmark scenarios, this includes
$U(1)^\prime_\psi$ models where the sneutrino mass lies in the 80-140 GeV
range. The situation is slightly better for neutralino LSP scenarios, where
several other viable $U(1)^\prime$ choices exist. The LSP mass in
these cases can either be small, when the neutralino LSP is mainly bino-like, or
much
larger, when the neutralino LSP has a large or dominant higgsino component and
is close in mass to the NLSP. This last configuration allows an
appropriate amount of co-annihilations with the NLSP and consequently avoids
any tension with the cosmological data.
We have moreover observed that in sneutrino LSP scenarios, the second lightest
Higgs boson is likely degenerate in mass with the $Z^\prime$ boson for
$U(1)^\prime_\eta$ and $U(1)^\prime_\psi$ models, which consists of an
indication that it is mostly singlet. In contrast, $U(1)^\prime_I$ scenarios
favour a lighter second Higgs boson with a larger doublet component.

Unfortunately these UMSSM scenarios do not have good prospects for observability
in present collider experiments, even when the high-luminosity run of the LHC is
considered. All signal cross sections are too small and the background is thus
overwhelming for all the possible associated new physics signatures. Our
predictions for the spin-independent cross section related to DM direct
detection are nonetheless within the reach of the future of XENON-1T
experiment~\cite{Aprile:2015uzo}, that is expected to be sensitive to cross
section values of about $1.6 \times 10^{-47}$ cm$^2$ for DM masses of 50-60~GeV.
This should allow for conclusive statements regarding the viability
of any of the UMSSM scenarios presented in this work.

\begin{acknowledgments}
The authors thank Gennaro Corcella and Florian Staub for their help with the
usage of the \textsc{Sarah} program, as well as Olivier Mattelaer and Torbj\"{o}rn
Sj\"ostrand for their help with \textsc{MadGraph5\_aMC@NLO} and \textsc{Pythia8}.
MF acknowledges NSERC for partial financial support under grant number SAP105354. The work of BF is partly supported by French state funds managed by the Agence Nationale de la Recherche (ANR), in the context of the LABEX ILP (ANR-11-IDEX-0004-02, ANR-10-LABX-63).
\end{acknowledgments}

\bibliography{umssm}

\providecommand{\href}[2]{#2}\begingroup\raggedright\begin{thebibliography}{10}

\bibitem{ATLAS:2016uzr}
{\scshape ATLAS} collaboration, \emph{{Search for pair production of gluinos
  decaying via top or bottom squarks in events with $b$-jets and large missing
  transverse momentum in $pp$ collisions at $\sqrt{s}=13$ TeV with the ATLAS
  detector}},  ATLAS-CONF-2016-052.

\bibitem{ATLAS:2016lsr}
{\scshape ATLAS} collaboration, \emph{{Search for squarks and gluinos in events
  with an isolated lepton, jets and missing transverse momentum at $\sqrt{s}$ =
  13 TeV with the ATLAS detector}},  ATLAS-CONF-2016-054.

\bibitem{ATLAS:2016kts}
{\scshape ATLAS} collaboration, \emph{{Further searches for squarks and gluinos
  in final states with jets and missing transverse momentum at $\sqrt{s}$ =13
  TeV with the ATLAS detector}},  ATLAS-CONF-2016-078.

\bibitem{Dine:2015xga}
M.~Dine, \emph{{Naturalness Under Stress}},
  \href{http://dx.doi.org/10.1146/annurev-nucl-102014-022053}{\emph{{Ann. Rev.
  Nucl. Part. Sci.}} {\bf 65} (2015) 43--62},
  [\href{http://arxiv.org/abs/1501.01035}{{\tt 1501.01035}}].

\bibitem{Hall:2011aa}
L.~J. Hall, D.~Pinner and J.~T. Ruderman, \emph{{A Natural SUSY Higgs Near 126
  GeV}}, \href{http://dx.doi.org/10.1007/JHEP04(2012)131}{\emph{JHEP} {\bf 04}
  (2012) 131}, [\href{http://arxiv.org/abs/1112.2703}{{\tt 1112.2703}}].

\bibitem{Aad:2012tfa}
{\scshape ATLAS} collaboration, G.~Aad et~al., \emph{Observation of a new
  particle in the search for the standard model higgs boson with the atlas
  detector at the lhc},
  \href{http://dx.doi.org/10.1016/j.physletb.2012.08.020}{\emph{Phys.Lett.}
  {\bf B716} (2012) 1--29}, [\href{http://arxiv.org/abs/1207.7214}{{\tt
  1207.7214}}].

\bibitem{Chatrchyan:2012xdj}
{\scshape CMS} collaboration, S.~Chatrchyan et~al., \emph{{Observation of a new
  boson at a mass of 125 GeV with the CMS experiment at the LHC}},
  \href{http://dx.doi.org/10.1016/j.physletb.2012.08.021}{\emph{Phys. Lett.}
  {\bf B716} (2012) 30--61}, [\href{http://arxiv.org/abs/1207.7235}{{\tt
  1207.7235}}].

\bibitem{Cassel:2011tg}
S.~Cassel, D.~M. Ghilencea, S.~Kraml, A.~Lessa and G.~G. Ross,
  \emph{{Fine-tuning implications for complementary dark matter and LHC SUSY
  searches}}, \href{http://dx.doi.org/10.1007/JHEP05(2011)120}{\emph{JHEP} {\bf
  05} (2011) 120}, [\href{http://arxiv.org/abs/1101.4664}{{\tt 1101.4664}}].

\bibitem{Baer:2014ica}
H.~Baer, V.~Barger, D.~Mickelson and M.~Padeffke-Kirkland, \emph{{SUSY models
  under siege: LHC constraints and electroweak fine-tuning}},
  \href{http://dx.doi.org/10.1103/PhysRevD.89.115019}{\emph{Phys. Rev.} {\bf
  D89} (2014) 115019}, [\href{http://arxiv.org/abs/1404.2277}{{\tt
  1404.2277}}].

\bibitem{Ellis:2010kf}
J.~Ellis and K.~A. Olive, \emph{{Supersymmetric Dark Matter Candidates}},
  \href{http://arxiv.org/abs/1001.3651}{{\tt 1001.3651}}.

\bibitem{Hewett:1988xc}
J.~L. Hewett and T.~G. Rizzo, \emph{{Low-Energy Phenomenology of Superstring
  Inspired E(6) Models}},
  \href{http://dx.doi.org/10.1016/0370-1573(89)90071-9}{\emph{Phys. Rept.} {\bf
  183} (1989) 193}.

\bibitem{Langacker:1998tc}
P.~Langacker and J.~Wang, \emph{{U(1)-prime symmetry breaking in supersymmetric
  E(6) models}},
  \href{http://dx.doi.org/10.1103/PhysRevD.58.115010}{\emph{Phys. Rev.} {\bf
  D58} (1998) 115010}, [\href{http://arxiv.org/abs/hep-ph/9804428}{{\tt
  hep-ph/9804428}}].

\bibitem{Hill:2002ap}
C.~T. Hill and E.~H. Simmons, \emph{{Strong dynamics and electroweak symmetry
  breaking}},
  \href{http://dx.doi.org/10.1016/S0370-1573(03)00140-6}{\emph{Phys. Rept.}
  {\bf 381} (2003) 235--402}, [\href{http://arxiv.org/abs/hep-ph/0203079}{{\tt
  hep-ph/0203079}}].

\bibitem{ArkaniHamed:2002qy}
N.~Arkani-Hamed, A.~G. Cohen, E.~Katz and A.~E. Nelson, \emph{{The Littlest
  Higgs}}, \href{http://dx.doi.org/10.1088/1126-6708/2002/07/034}{\emph{JHEP}
  {\bf 07} (2002) 034}, [\href{http://arxiv.org/abs/hep-ph/0206021}{{\tt
  hep-ph/0206021}}].

\bibitem{Han:2003wu}
T.~Han, H.~E. Logan, B.~McElrath and L.-T. Wang, \emph{{Phenomenology of the
  little Higgs model}},
  \href{http://dx.doi.org/10.1103/PhysRevD.67.095004}{\emph{Phys. Rev.} {\bf
  D67} (2003) 095004}, [\href{http://arxiv.org/abs/hep-ph/0301040}{{\tt
  hep-ph/0301040}}].

\bibitem{Antoniadis:1990ew}
I.~Antoniadis, \emph{{A Possible new dimension at a few TeV}},
  \href{http://dx.doi.org/10.1016/0370-2693(90)90617-F}{\emph{Phys. Lett.} {\bf
  B246} (1990) 377--384}.

\bibitem{Cvetic:1997ky}
M.~Cvetic, D.~A. Demir, J.~R. Espinosa, L.~L. Everett and P.~Langacker,
  \emph{{Electroweak breaking and the mu problem in supergravity models with an
  additional U(1)}}, \href{http://dx.doi.org/10.1103/PhysRevD.56.2861,
  10.1103/PhysRevD.58.119905}{\emph{Phys. Rev.} {\bf D56} (1997) 2861},
  [\href{http://arxiv.org/abs/hep-ph/9703317}{{\tt hep-ph/9703317}}].

\bibitem{Demir:2005ti}
D.~A. Demir, G.~L. Kane and T.~T. Wang, \emph{{The Minimal U(1)' extension of
  the MSSM}}, \href{http://dx.doi.org/10.1103/PhysRevD.72.015012}{\emph{Phys.
  Rev.} {\bf D72} (2005) 015012},
  [\href{http://arxiv.org/abs/hep-ph/0503290}{{\tt hep-ph/0503290}}].

\bibitem{Ellwanger:2009dp}
U.~Ellwanger, C.~Hugonie and A.~M. Teixeira, \emph{{The Next-to-Minimal
  Supersymmetric Standard Model}},
  \href{http://dx.doi.org/10.1016/j.physrep.2010.07.001}{\emph{Phys. Rept.}
  {\bf 496} (2010) 1--77}, [\href{http://arxiv.org/abs/0910.1785}{{\tt
  0910.1785}}].

\bibitem{Ellis:1986mq}
J.~R. Ellis, K.~Enqvist, D.~V. Nanopoulos, K.~A. Olive, M.~Quiros and
  F.~Zwirner, \emph{{Problems for (2,0) Compactifications}},
  \href{http://dx.doi.org/10.1016/0370-2693(86)90185-1}{\emph{Phys. Lett.} {\bf
  B176} (1986) 403--408}.

\bibitem{Langacker:2000ju}
P.~Langacker and M.~Plumacher, \emph{{Flavor changing effects in theories with
  a heavy $Z^\prime$ boson with family nonuniversal couplings}},
  \href{http://dx.doi.org/10.1103/PhysRevD.62.013006}{\emph{Phys. Rev.} {\bf
  D62} (2000) 013006}, [\href{http://arxiv.org/abs/hep-ph/0001204}{{\tt
  hep-ph/0001204}}].

\bibitem{Hinshaw:2012aka}
{\scshape WMAP} collaboration, G.~Hinshaw et~al., \emph{{Nine-Year Wilkinson
  Microwave Anisotropy Probe (WMAP) Observations: Cosmological Parameter
  Results}},
  \href{http://dx.doi.org/10.1088/0067-0049/208/2/19}{\emph{Astrophys. J.
  Suppl.} {\bf 208} (2013) 19}, [\href{http://arxiv.org/abs/1212.5226}{{\tt
  1212.5226}}].

\bibitem{Ade:2013zuv}
{\scshape Planck} collaboration, P.~A.~R. Ade et~al., \emph{{Planck 2013
  results. XVI. Cosmological parameters}},
  \href{http://dx.doi.org/10.1051/0004-6361/201321591}{\emph{Astron.
  Astrophys.} {\bf 571} (2014) A16},
  [\href{http://arxiv.org/abs/1303.5076}{{\tt 1303.5076}}].

\bibitem{Ma:1986we}
E.~Ma, \emph{{Particle Dichotomy and Left-Right Decomposition of E(6)
  Superstring Models}},
  \href{http://dx.doi.org/10.1103/PhysRevD.36.274}{\emph{Phys. Rev.} {\bf D36}
  (1987) 274}.

\bibitem{Ham:2007kc}
S.~W. Ham, E.~J. Yoo and S.~K. Oh, \emph{{Explicit CP violation in a MSSM with
  an extra U(1)-prime}},
  \href{http://dx.doi.org/10.1103/PhysRevD.76.015004}{\emph{Phys. Rev.} {\bf
  D76} (2007) 015004}, [\href{http://arxiv.org/abs/hep-ph/0703041}{{\tt
  hep-ph/0703041}}].

\bibitem{Langacker:2008yv}
P.~Langacker, \emph{{The Physics of Heavy $Z^\prime$ Gauge Bosons}},
  \href{http://dx.doi.org/10.1103/RevModPhys.81.1199}{\emph{Rev. Mod. Phys.}
  {\bf 81} (2009) 1199--1228}, [\href{http://arxiv.org/abs/0801.1345}{{\tt
  0801.1345}}].

\bibitem{Belanger:2015cra}
G.~B{\'e}langer, J.~Da~Silva, U.~Laa and A.~Pukhov, \emph{{Probing U(1)
  extensions of the MSSM at the LHC Run I and in dark matter searches}},
  \href{http://dx.doi.org/10.1007/JHEP09(2015)151}{\emph{JHEP} {\bf 09} (2015)
  151}, [\href{http://arxiv.org/abs/1505.06243}{{\tt 1505.06243}}].

\bibitem{Belanger:2011rs}
G.~Belanger, J.~Da~Silva and A.~Pukhov, \emph{{The Right-handed sneutrino as
  thermal dark matter in U(1) extensions of the MSSM}},
  \href{http://dx.doi.org/10.1088/1475-7516/2011/12/014}{\emph{JCAP} {\bf 1112}
  (2011) 014}, [\href{http://arxiv.org/abs/1110.2414}{{\tt 1110.2414}}].

\bibitem{Corcella:2014lha}
G.~Corcella, \emph{{Phenomenology of supersymmetric $ Z^\prime $ decays at the
  Large Hadron Collider}},
  \href{http://dx.doi.org/10.1140/epjc/s10052-015-3459-9}{\emph{Eur. Phys. J.}
  {\bf C75} (2015) 264}, [\href{http://arxiv.org/abs/1412.6831}{{\tt
  1412.6831}}].

\bibitem{Corcella:2012dw}
G.~Corcella and S.~Gentile, \emph{{Heavy Neutral Gauge Bosons at the LHC in an
  Extended MSSM}}, \href{http://dx.doi.org/10.1016/j.nuclphysb.2012.09.009,
  10.1016/j.nuclphysb.2012.11.024}{\emph{Nucl. Phys.} {\bf B866} (2013)
  293--336}, [\href{http://arxiv.org/abs/1205.5780}{{\tt 1205.5780}}].

\bibitem{BhupalDev:2012ru}
P.~S. Bhupal~Dev, S.~Mondal, B.~Mukhopadhyaya and S.~Roy, \emph{{Phenomenology
  of Light Sneutrino Dark Matter in cMSSM/mSUGRA with Inverse Seesaw}},
  \href{http://dx.doi.org/10.1007/JHEP09(2012)110}{\emph{JHEP} {\bf 09} (2012)
  110}, [\href{http://arxiv.org/abs/1207.6542}{{\tt 1207.6542}}].

\bibitem{Barger:2012ey}
V.~Barger, D.~Marfatia and A.~Peterson, \emph{{LHC and dark matter signals of
  $Z'$ bosons}},
  \href{http://dx.doi.org/10.1103/PhysRevD.87.015026}{\emph{Phys. Rev.} {\bf
  D87} (2013) 015026}, [\href{http://arxiv.org/abs/1206.6649}{{\tt
  1206.6649}}].

\bibitem{Chiang:2014yva}
C.-W. Chiang, T.~Nomura and K.~Yagyu, \emph{{Phenomenology of $E_6$-Inspired
  Leptophobic $Z'$ Boson at the LHC}},
  \href{http://dx.doi.org/10.1007/JHEP05(2014)106}{\emph{JHEP} {\bf 05} (2014)
  106}, [\href{http://arxiv.org/abs/1402.5579}{{\tt 1402.5579}}].

\bibitem{Belanger:2017vpq}
G.~Bélanger, J.~Da~Silva and H.~M. Tran, \emph{{Dark matter in U(1) extensions
  of the MSSM with gauge kinetic mixing}},
  \href{http://arxiv.org/abs/1703.03275}{{\tt 1703.03275}}.

\bibitem{Erler:2000wu}
J.~Erler, \emph{{Chiral models of weak scale supersymmetry}},
  \href{http://dx.doi.org/10.1016/S0550-3213(00)00427-2}{\emph{Nucl. Phys.}
  {\bf B586} (2000) 73--91}, [\href{http://arxiv.org/abs/hep-ph/0006051}{{\tt
  hep-ph/0006051}}].

\bibitem{Deppisch:2007xu}
F.~Deppisch, A.~Freitas, W.~Porod and P.~M. Zerwas, \emph{{Determining Heavy
  Mass Parameters in Supersymmetric SO(10) Models}},
  \href{http://dx.doi.org/10.1103/PhysRevD.77.075009}{\emph{Phys. Rev.} {\bf
  D77} (2008) 075009}, [\href{http://arxiv.org/abs/0712.0361}{{\tt
  0712.0361}}].

\bibitem{Kim:1983dt}
J.~E. Kim and H.~P. Nilles, \emph{{The mu Problem and the Strong CP Problem}},
  \href{http://dx.doi.org/10.1016/0370-2693(84)91890-2}{\emph{Phys. Lett.} {\bf
  B138} (1984) 150}.

\bibitem{Suematsu:1994qm}
D.~Suematsu and Y.~Yamagishi, \emph{{Radiative symmetry breaking in a
  supersymmetric model with an extra U(1)}},
  \href{http://dx.doi.org/10.1142/S0217751X95002096}{\emph{Int. J. Mod. Phys.}
  {\bf A10} (1995) 4521--4536},
  [\href{http://arxiv.org/abs/hep-ph/9411239}{{\tt hep-ph/9411239}}].

\bibitem{Cvetic:1996mf}
M.~Cvetic and P.~Langacker, \emph{{New gauge bosons from string models}},
  \href{http://dx.doi.org/10.1142/S0217732396001260}{\emph{Mod. Phys. Lett.}
  {\bf A11} (1996) 1247--1262},
  [\href{http://arxiv.org/abs/hep-ph/9602424}{{\tt hep-ph/9602424}}].

\bibitem{Jain:1995cb}
V.~Jain and R.~Shrock, \emph{{U(1)-A models of fermion masses without a mu
  problem}},  \href{http://arxiv.org/abs/hep-ph/9507238}{{\tt hep-ph/9507238}}.

\bibitem{Demir:1998dm}
D.~A. Demir, \emph{{Two Higgs doublet models from TeV scale supersymmetric
  extra U(1) models}},
  \href{http://dx.doi.org/10.1103/PhysRevD.59.015002}{\emph{Phys. Rev.} {\bf
  D59} (1999) 015002}, [\href{http://arxiv.org/abs/hep-ph/9809358}{{\tt
  hep-ph/9809358}}].

\bibitem{Minkowski:1977sc}
P.~Minkowski, \emph{{$\mu \to e\gamma$ at a Rate of One Out of $10^{9}$ Muon
  Decays?}}, \href{http://dx.doi.org/10.1016/0370-2693(77)90435-X}{\emph{Phys.
  Lett.} {\bf B67} (1977) 421--428}.

\bibitem{Mohapatra:1979ia}
R.~N. Mohapatra and G.~Senjanovic, \emph{{Neutrino Mass and Spontaneous Parity
  Violation}}, \href{http://dx.doi.org/10.1103/PhysRevLett.44.912}{\emph{Phys.
  Rev. Lett.} {\bf 44} (1980) 912}.

\bibitem{Schechter:1980gr}
J.~Schechter and J.~W.~F. Valle, \emph{{Neutrino Masses in SU(2) x U(1)
  Theories}}, \href{http://dx.doi.org/10.1103/PhysRevD.22.2227}{\emph{Phys.
  Rev.} {\bf D22} (1980) 2227}.

\bibitem{Schechter:1981cv}
J.~Schechter and J.~W.~F. Valle, \emph{{Neutrino Decay and Spontaneous
  Violation of Lepton Number}},
  \href{http://dx.doi.org/10.1103/PhysRevD.25.774}{\emph{Phys. Rev.} {\bf D25}
  (1982) 774}.

\bibitem{Kang:2004ix}
J.-h. Kang, P.~Langacker and T.-j. Li, \emph{{Neutrino masses in supersymmetric
  SU(3)(C) x SU(2)(L) x U(1)(Y) x U(1)-prime models}},
  \href{http://dx.doi.org/10.1103/PhysRevD.71.015012}{\emph{Phys. Rev.} {\bf
  D71} (2005) 015012}, [\href{http://arxiv.org/abs/hep-ph/0411404}{{\tt
  hep-ph/0411404}}].

\bibitem{Demir:2006jj}
D.~A. Demir and Y.~Farzan, \emph{{Correlating mu parameter and right-handed
  neutrino masses in N=1 supergravity}},
  \href{http://dx.doi.org/10.1088/1126-6708/2006/03/010}{\emph{JHEP} {\bf 03}
  (2006) 010}, [\href{http://arxiv.org/abs/hep-ph/0601096}{{\tt
  hep-ph/0601096}}].

\bibitem{Demir:2007dt}
D.~A. Demir, L.~L. Everett and P.~Langacker, \emph{{Dirac Neutrino Masses from
  Generalized Supersymmetry Breaking}},
  \href{http://dx.doi.org/10.1103/PhysRevLett.100.091804}{\emph{Phys. Rev.
  Lett.} {\bf 100} (2008) 091804}, [\href{http://arxiv.org/abs/0712.1341}{{\tt
  0712.1341}}].

\bibitem{Heinemeyer:2011aa}
S.~Heinemeyer, O.~Stal and G.~Weiglein, \emph{{Interpreting the LHC Higgs
  Search Results in the MSSM}},
  \href{http://dx.doi.org/10.1016/j.physletb.2012.02.084}{\emph{Phys. Lett.}
  {\bf B710} (2012) 201--206}, [\href{http://arxiv.org/abs/1112.3026}{{\tt
  1112.3026}}].

\bibitem{Ross:2012nr}
G.~G. Ross, K.~Schmidt-Hoberg and F.~Staub, \emph{{The Generalised NMSSM at One
  Loop: Fine Tuning and Phenomenology}},
  \href{http://dx.doi.org/10.1007/JHEP08(2012)074}{\emph{JHEP} {\bf 08} (2012)
  074}, [\href{http://arxiv.org/abs/1205.1509}{{\tt 1205.1509}}].

\bibitem{Staub:2013tta}
F.~Staub, \emph{{SARAH 4 : A tool for (not only SUSY) model builders}},
  \href{http://dx.doi.org/10.1016/j.cpc.2014.02.018}{\emph{Comput. Phys.
  Commun.} {\bf 185} (2014) 1773--1790},
  [\href{http://arxiv.org/abs/1309.7223}{{\tt 1309.7223}}].

\bibitem{Porod:2011nf}
W.~Porod and F.~Staub, \emph{{SPheno 3.1: Extensions including flavour,
  CP-phases and models beyond the MSSM}},
  \href{http://dx.doi.org/10.1016/j.cpc.2012.05.021}{\emph{Comput. Phys.
  Commun.} {\bf 183} (2012) 2458--2469},
  [\href{http://arxiv.org/abs/1104.1573}{{\tt 1104.1573}}].

\bibitem{Belanger:2014vza}
G.~B{\'e}langer, F.~Boudjema, A.~Pukhov and A.~Semenov, \emph{{micrOMEGAs4.1:
  two dark matter candidates}},
  \href{http://dx.doi.org/10.1016/j.cpc.2015.03.003}{\emph{Comput. Phys.
  Commun.} {\bf 192} (2015) 322--329},
  [\href{http://arxiv.org/abs/1407.6129}{{\tt 1407.6129}}].

\bibitem{Bechtle:2008jh}
P.~Bechtle, O.~Brein, S.~Heinemeyer, G.~Weiglein and K.~E. Williams,
  \emph{{HiggsBounds: Confronting Arbitrary Higgs Sectors with Exclusion Bounds
  from LEP and the Tevatron}},
  \href{http://dx.doi.org/10.1016/j.cpc.2009.09.003}{\emph{Comput. Phys.
  Commun.} {\bf 181} (2010) 138--167},
  [\href{http://arxiv.org/abs/0811.4169}{{\tt 0811.4169}}].

\bibitem{Bechtle:2013xfa}
P.~Bechtle, S.~Heinemeyer, O.~St{\aa}l, T.~Stefaniak and G.~Weiglein,
  \emph{{$HiggsSignals$: Confronting arbitrary Higgs sectors with measurements
  at the Tevatron and the LHC}},
  \href{http://dx.doi.org/10.1140/epjc/s10052-013-2711-4}{\emph{Eur. Phys. J.}
  {\bf C74} (2014) 2711}, [\href{http://arxiv.org/abs/1305.1933}{{\tt
  1305.1933}}].

\bibitem{Buckley:2013jua}
A.~Buckley, \emph{{PySLHA: a Pythonic interface to SUSY Les Houches Accord
  data}}, \href{http://dx.doi.org/10.1140/epjc/s10052-015-3638-8}{\emph{Eur.
  Phys. J.} {\bf C75} (2015) 467}, [\href{http://arxiv.org/abs/1305.4194}{{\tt
  1305.4194}}].

\bibitem{Erler:2009jh}
J.~Erler, P.~Langacker, S.~Munir and E.~Rojas, \emph{{Improved Constraints on
  Z-prime Bosons from Electroweak Precision Data}},
  \href{http://dx.doi.org/10.1088/1126-6708/2009/08/017}{\emph{JHEP} {\bf 08}
  (2009) 017}, [\href{http://arxiv.org/abs/0906.2435}{{\tt 0906.2435}}].

\bibitem{Khachatryan:2016xvy}
{\scshape CMS} collaboration, V.~Khachatryan et~al., \emph{{Search for new
  physics with the M$_{T2}$ variable in all-jets final states produced in pp
  collisions at $ \sqrt{s}=13 $ TeV}},
  \href{http://dx.doi.org/10.1007/JHEP10(2016)006}{\emph{JHEP} {\bf 10} (2016)
  006}, [\href{http://arxiv.org/abs/1603.04053}{{\tt 1603.04053}}].

\bibitem{Olive:2016xmw}
{\scshape Particle Data Group} collaboration, C.~Patrignani et~al.,
  \emph{{Review of Particle Physics}},
  \href{http://dx.doi.org/10.1088/1674-1137/40/10/100001}{\emph{Chin. Phys.}
  {\bf C40} (2016) 100001}.

\bibitem{Aaboud:2016lwz}
{\scshape ATLAS} collaboration, M.~Aaboud et~al., \emph{{Search for top squarks
  in final states with one isolated lepton, jets, and missing transverse
  momentum in $\sqrt{s}=13$ TeV $pp$ collisions with the ATLAS detector}},
  \href{http://dx.doi.org/10.1103/PhysRevD.94.052009}{\emph{Phys. Rev.} {\bf
  D94} (2016) 052009}, [\href{http://arxiv.org/abs/1606.03903}{{\tt
  1606.03903}}].

\bibitem{Aaij:2012nna}
{\scshape LHCb} collaboration, R.~Aaij et~al., \emph{{First Evidence for the
  Decay $B_s^0 \to \mu^+ \mu^-$}},
  \href{http://dx.doi.org/10.1103/PhysRevLett.110.021801}{\emph{Phys. Rev.
  Lett.} {\bf 110} (2013) 021801}, [\href{http://arxiv.org/abs/1211.2674}{{\tt
  1211.2674}}].

\bibitem{Asner:2010qj}
{\scshape Heavy Flavor Averaging Group} collaboration, D.~Asner et~al.,
  \emph{{Averages of $b$-hadron, $c$-hadron, and $\tau$-lepton properties}},
  \href{http://arxiv.org/abs/1010.1589}{{\tt 1010.1589}}.

\bibitem{Amhis:2016xyh}
Y.~Amhis et~al., \emph{{Averages of $b$-hadron, $c$-hadron, and $\tau$-lepton
  properties as of summer 2016}},  \href{http://arxiv.org/abs/1612.07233}{{\tt
  1612.07233}}.

\bibitem{Corcella:2013cma}
G.~Corcella, \emph{{Searching for supersymmetry in $Z'$ decays}},
  \href{http://dx.doi.org/10.1051/epjconf/20136018011}{\emph{EPJ Web Conf.}
  {\bf 60} (2013) 18011}, [\href{http://arxiv.org/abs/1307.1040}{{\tt
  1307.1040}}].

\bibitem{Harlander:2012pb}
R.~V. Harlander, S.~Liebler and H.~Mantler, \emph{{SusHi: A program for the
  calculation of Higgs production in gluon fusion and bottom-quark annihilation
  in the Standard Model and the MSSM}},
  \href{http://dx.doi.org/10.1016/j.cpc.2013.02.006}{\emph{Comput. Phys.
  Commun.} {\bf 184} (2013) 1605--1617},
  [\href{http://arxiv.org/abs/1212.3249}{{\tt 1212.3249}}].

\bibitem{Heinemeyer:2013tqa}
{\scshape LHC Higgs Cross Section Working Group} collaboration, J.~R. Andersen
  et~al., \emph{{Handbook of LHC Higgs Cross Sections: 3. Higgs Properties}},
  \href{http://arxiv.org/abs/1307.1347}{{\tt 1307.1347}}.

\bibitem{Bennett:2006fi}
{\scshape Muon g-2} collaboration, G.~W. Bennett et~al., \emph{{Final Report of
  the Muon E821 Anomalous Magnetic Moment Measurement at BNL}},
  \href{http://dx.doi.org/10.1103/PhysRevD.73.072003}{\emph{Phys. Rev.} {\bf
  D73} (2006) 072003}, [\href{http://arxiv.org/abs/hep-ex/0602035}{{\tt
  hep-ex/0602035}}].

\bibitem{Grange:2015fou}
{\scshape Muon g-2} collaboration, J.~Grange et~al., \emph{{Muon (g-2)
  Technical Design Report}},  \href{http://arxiv.org/abs/1501.06858}{{\tt
  1501.06858}}.

\bibitem{Saito:2012zz}
{\scshape J-PARC g-'2/EDM} collaboration, N.~Saito, \emph{{A novel precision
  measurement of muon g-2 and EDM at J-PARC}},
  \href{http://dx.doi.org/10.1063/1.4742078}{\emph{AIP Conf. Proc.} {\bf 1467}
  (2012) 45--56}.

\bibitem{Komatsu:2010fb}
{\scshape WMAP} collaboration, E.~Komatsu et~al., \emph{{Seven-Year Wilkinson
  Microwave Anisotropy Probe (WMAP) Observations: Cosmological
  Interpretation}},
  \href{http://dx.doi.org/10.1088/0067-0049/192/2/18}{\emph{Astrophys. J.
  Suppl.} {\bf 192} (2011) 18}, [\href{http://arxiv.org/abs/1001.4538}{{\tt
  1001.4538}}].

\bibitem{Spergel:2006hy}
{\scshape WMAP} collaboration, D.~N. Spergel et~al., \emph{{Wilkinson Microwave
  Anisotropy Probe (WMAP) three year results: implications for cosmology}},
  \href{http://dx.doi.org/10.1086/513700}{\emph{Astrophys. J. Suppl.} {\bf 170}
  (2007) 377}, [\href{http://arxiv.org/abs/astro-ph/0603449}{{\tt
  astro-ph/0603449}}].

\bibitem{Akerib:2015rjg}
{\scshape LUX} collaboration, D.~S. Akerib et~al., \emph{{Improved Limits on
  Scattering of Weakly Interacting Massive Particles from Reanalysis of 2013
  LUX Data}},
  \href{http://dx.doi.org/10.1103/PhysRevLett.116.161301}{\emph{Phys. Rev.
  Lett.} {\bf 116} (2016) 161301}, [\href{http://arxiv.org/abs/1512.03506}{{\tt
  1512.03506}}].

\bibitem{mark_mitchell_2016_61108}
M.~Mitchell, B.~Muftakhidinov, T.~Winchen, Z.~Jedrzejewski-Szmek and T.~G.
  Badger, \emph{engauge-digitizer: Support for smaller monitors},  Aug., 2016.
\newblock 10.5281/zenodo.61108.

\bibitem{Abbasi:2009uz}
{\scshape IceCube} collaboration, R.~Abbasi et~al., \emph{{Limits on a muon
  flux from neutralino annihilations in the Sun with the IceCube 22-string
  detector}},
  \href{http://dx.doi.org/10.1103/PhysRevLett.102.201302}{\emph{Phys. Rev.
  Lett.} {\bf 102} (2009) 201302}, [\href{http://arxiv.org/abs/0902.2460}{{\tt
  0902.2460}}].

\bibitem{Baer:2011ab}
H.~Baer, V.~Barger and A.~Mustafayev, \emph{{Implications of a 125 GeV Higgs
  scalar for LHC SUSY and neutralino dark matter searches}},
  \href{http://dx.doi.org/10.1103/PhysRevD.85.075010}{\emph{Phys. Rev.} {\bf
  D85} (2012) 075010}, [\href{http://arxiv.org/abs/1112.3017}{{\tt
  1112.3017}}].

\bibitem{Arina:2016cqj}
C.~Arina et~al., \emph{{A comprehensive approach to dark matter studies:
  exploration of simplified top-philic models}},
  \href{http://dx.doi.org/10.1007/JHEP11(2016)111}{\emph{JHEP} {\bf 11} (2016)
  111}, [\href{http://arxiv.org/abs/1605.09242}{{\tt 1605.09242}}].

\bibitem{Degrande:2011ua}
C.~Degrande, C.~Duhr, B.~Fuks, D.~Grellscheid, O.~Mattelaer and T.~Reiter,
  \emph{{UFO - The Universal FeynRules Output}},
  \href{http://dx.doi.org/10.1016/j.cpc.2012.01.022}{\emph{Comput. Phys.
  Commun.} {\bf 183} (2012) 1201--1214},
  [\href{http://arxiv.org/abs/1108.2040}{{\tt 1108.2040}}].

\bibitem{Alwall:2014hca}
J.~Alwall, R.~Frederix, S.~Frixione, V.~Hirschi, F.~Maltoni, O.~Mattelaer
  et~al., \emph{{The automated computation of tree-level and next-to-leading
  order differential cross sections, and their matching to parton shower
  simulations}}, \href{http://dx.doi.org/10.1007/JHEP07(2014)079}{\emph{JHEP}
  {\bf 07} (2014) 079}, [\href{http://arxiv.org/abs/1405.0301}{{\tt
  1405.0301}}].

\bibitem{Sjostrand:2014zea}
T.~Sjöstrand, S.~Ask, J.~R. Christiansen, R.~Corke, N.~Desai, P.~Ilten et~al.,
  \emph{{An Introduction to PYTHIA 8.2}},
  \href{http://dx.doi.org/10.1016/j.cpc.2015.01.024}{\emph{Comput. Phys.
  Commun.} {\bf 191} (2015) 159--177},
  [\href{http://arxiv.org/abs/1410.3012}{{\tt 1410.3012}}].

\bibitem{deFavereau:2013fsa}
{\scshape DELPHES 3} collaboration, J.~de~Favereau, C.~Delaere, P.~Demin,
  A.~Giammanco, V.~Lemaître, A.~Mertens et~al., \emph{{DELPHES 3, A modular
  framework for fast simulation of a generic collider experiment}},
  \href{http://dx.doi.org/10.1007/JHEP02(2014)057}{\emph{JHEP} {\bf 02} (2014)
  057}, [\href{http://arxiv.org/abs/1307.6346}{{\tt 1307.6346}}].

\bibitem{Cacciari:2008gp}
M.~Cacciari, G.~P. Salam and G.~Soyez, \emph{{The Anti-k(t) jet clustering
  algorithm}},
  \href{http://dx.doi.org/10.1088/1126-6708/2008/04/063}{\emph{JHEP} {\bf 04}
  (2008) 063}, [\href{http://arxiv.org/abs/0802.1189}{{\tt 0802.1189}}].

\bibitem{Cacciari:2011ma}
M.~Cacciari, G.~P. Salam and G.~Soyez, \emph{{FastJet User Manual}},
  \href{http://dx.doi.org/10.1140/epjc/s10052-012-1896-2}{\emph{Eur. Phys. J.}
  {\bf C72} (2012) 1896}, [\href{http://arxiv.org/abs/1111.6097}{{\tt
  1111.6097}}].

\bibitem{Conte:2012fm}
E.~Conte, B.~Fuks and G.~Serret, \emph{{MadAnalysis 5, A User-Friendly
  Framework for Collider Phenomenology}},
  \href{http://dx.doi.org/10.1016/j.cpc.2012.09.009}{\emph{Comput. Phys.
  Commun.} {\bf 184} (2013) 222--256},
  [\href{http://arxiv.org/abs/1206.1599}{{\tt 1206.1599}}].

\bibitem{Frank:2012ne}
M.~Frank, L.~Selbuz and I.~Turan, \emph{{Neutralino and Chargino Production in
  U(1)' at the LHC}},
  \href{http://dx.doi.org/10.1140/epjc/s10052-013-2656-7}{\emph{Eur. Phys. J.}
  {\bf C73} (2013) 2656}, [\href{http://arxiv.org/abs/1212.4428}{{\tt
  1212.4428}}].

\bibitem{Aprile:2015uzo}
{\scshape XENON} collaboration, E.~Aprile et~al., \emph{{Physics reach of the
  XENON1T dark matter experiment}},
  \href{http://dx.doi.org/10.1088/1475-7516/2016/04/027}{\emph{JCAP} {\bf 1604}
  (2016) 027}, [\href{http://arxiv.org/abs/1512.07501}{{\tt 1512.07501}}].

\end{thebibliography}\endgroup

\end{document}